\documentclass[11pt,reqno]{article}
\usepackage{bm}
\usepackage{amsthm}
\usepackage{amsmath}
\usepackage{amssymb}
\usepackage{pdfsync}
\usepackage[round]{natbib}
\usepackage{color}
\usepackage{authblk}
\usepackage{microtype}
\usepackage{graphicx}
\usepackage{float}
\usepackage{prodint}
\usepackage{paralist}
\usepackage{tabularx}
\usepackage{epstopdf}
\usepackage{rotating}
\usepackage{geometry}
\DeclareMathSymbol{\shortminus}{\mathbin}{AMSa}{"39}
\newcommand{\rnc}{\renewcommand}
\newcommand{\nc}{\newcommand}

\renewcommand{\hat}{\widehat}
\nc{\mb}{\mathbb}
\nc{\mc}{\mathcal}
\nc{\N}{\mb{N}}
\nc{\R}{\mb{R}}
\nc{\Q}{\mb{Q}}

\nc{\E}{E}
\rnc{\P}{P}
\nc{\var}{ \text{Var} }
\nc{\mbf}{\boldsymbol}
\nc{\I}{I}

\nc{\trans}{^{\top}}
\nc{\assumption}{{Assumption A}}

\newcommand\blfootnote[1]{%
  \begingroup
  \renewcommand\thefootnote{}\footnote{#1}%
  \addtocounter{footnote}{-1}%
  \endgroup
} % Fuer Fussnoten ohne Nummer

\DeclareSymbolFont{matha}{OML}{txmi}{m}{it}% txfonts
\DeclareMathSymbol{\varv}{\mathord}{matha}{118}

\usepackage{url}
\usepackage{longtable}
\usepackage{booktabs}

\usepackage{caption}
\DeclareCaptionListFormat{ignore}{}

\begin{document}

\title{\Large \bf
Functional repeated measures analysis of variance and its application}
\author[1]{Katarzyna Kury\l o}
\author[1,$^*$]{\L ukasz Smaga}

\affil[1]{Faculty of Mathematics and Computer Science, Adam Mickiewicz University, Poland}

\maketitle

\begin{abstract}
This paper is motivated by medical studies in which the same patients with multiple sclerosis are examined at several successive visits and described by fractional anisotropy tract profiles, which can be represented as functions. Since the observations for each patient are dependent random processes, they follow a repeated measures design for functional data. To compare the results for different visits, we thus consider functional repeated measures analysis of variance. For this purpose, a pointwise test statistic is constructed by adapting the classical test statistic for one-way repeated measures analysis of variance to the functional data framework. By integrating and taking the supremum of the pointwise test statistic, we create two global test statistics. Apart from verifying the general null hypothesis on the equality of mean functions corresponding to different objects, we also propose a simple method for post hoc analysis. We illustrate the finite sample properties of permutation and bootstrap testing procedures in an extensive simulation study. Finally, we analyze a motivating real data example in detail.
	
\blfootnote{${}^*$ Corresponding author. Email address: {ls@amu.edu.pl}}
\end{abstract}


%\vfill
%\vfill

\section{Introduction}
\label{sec_1}
Functional data analysis (FDA) is a branch of statistics which analyzes observations treated as functions, curves, or surfaces. To represent the data in such a way, one needs only to measure some variable over time or space, which is a scenario encountered in many fields. Then the discrete data observed at so-called design time points can be transformed into functional data. Such a representation allows us to avoid many problems of classical multivariate statistical methods, for example, the curse of dimensionality and missing data. Therefore, in the last two decades, numerous methods have been developed for classification, clustering, dimension reduction, regression, and statistical hypothesis testing for functional data. We refer to the following key textbooks in FDA for methodology, real data examples, and computational aspects of these methods: Ferraty and Vieu (2006), Horv\'ath and Kokoszka (2012), Ramsay and Silverman (2005), Ramsay et al. (2009), and Zhang (2013).

In the literature and in software, many tests can be found for standard functional analysis of variance (FANOVA), where the samples corresponding to different objects are independent (see, for example, Cuevas et al., 2004; Mrkvi\v{c}ka et al., 2020; Pini et al., 2018; Smaga and Zhang, 2020; Zhang, 2013; Zhang et al., 2019 and references therein). However, the independence assumption does not always hold for functional samples. In this paper, we consider a motivating real data example where the same patients with multiple sclerosis are examined using the magnetic resonance imaging technique at several successive visits. For each visit, the results of the examination are curves observed at some design time points, which allows them to be treated as functional data. It is of interest to determine whether or not the examined functional variable changes at the different visits. Examples of similar problems appear in many fields, including medicine, education, the social sciences, and psychology. Due to the dependence of functional samples corresponding to different visits, we have to consider repeated measures analysis of variance (ANOVA) in the framework of functional data. In this case, the literature is not so rich as for standard FANOVA. The repeated measures ANOVA problem for functional data was apparently first considered by Mart\'inez-Camblor and Corral (2011). Developing an idea of Cuevas et al. (2004), they proposed bootstrap and permutation tests based on the integral of the square of differences between sample mean functions. In the case of two groups, Smaga (2019) constructed a test based on the same test statistic, but to approximate its null distribution he used the Box-type approximation method (Box, 1954). This approach resulted in a testing procedure which is much less time-consuming than the tests of Mart\'inez-Camblor and Corral (2011), while being comparable to them in terms of type I error level control and power. Note that these statistical tests do not take into account information about variance (see Section~\ref{sec_2} for details). This is not an isolated case; for example, the $L^2$-norm-based test statistic for the standard FANOVA problem is also constructed based on the distance between sample mean functions only (Zhang, 2013). However, using information about variance can improve the test, as was shown in particular by Smaga (2020) for the paired two-sample problem for functional data. In a short note, Smaga (2021) extended these results from the two-sample case to the strict ANOVA case, i.e., when the number of groups is greater than two. In the present paper, the ideas presented in that note are developed in greater detail. First, we present the methodology at length, showing how the construction of tests for the global null hypothesis for functional data is based on classic ANOVA for real data. Moreover, we propose a simple post hoc method, which appears to be powerful (see Sections~\ref{sec_4} and~\ref{sec_5}). To obtain tests with good finite sample properties even for smaller sample sizes, we investigate different bootstrap and permutation procedures. These demonstrate good qualities in many models (see, for example, Amro et al., 2021; Ditzhaus et al., 2021; Konietschke et al., 2015; Konietschke and Pauly, 2014; Smaga and Zhang, 2020). However, this is not an invariable rule, as we also note in the extensive simulation study, where we check the control of type I error level and power of the considered tests in comparison with the tests of Mart\'inez-Camblor and Corral (2011) and their modifications. The simulation study is conducted for artificial and real data. The latter case explains the results of the hypothesis testing for a diffusion tensor imaging data set. This real data example demonstrates the applicability of the presented methods in an important problem.

The remainder of the paper is organized as follows: In Section~\ref{sec_2} we recall the repeated measures ANOVA model for random variables and present its counterpart for functional data. Moreover, we consider different pointwise test statistics for given statistical hypotheses. Section~\ref{sec_3} presents the global test statistics, permutation and bootstrap tests based on them, and the simple post hoc procedure for a functional repeated measures ANOVA problem. The type I error level control and power of these testing procedures are investigated in the simulation study in Section~\ref{sec_4}. In Section~\ref{sec_5}, we present a real data application for diffusion tensor imaging data. Finally, Section~\ref{sec_6} concludes the paper.

\section{Functional repeated measures ANOVA model and pointwise test statistics}
\label{sec_2}
In this section, we first recall the repeated measures ANOVA problem for real data, and then we present the counterpart of this problem for functional data. Finally, we define new pointwise test statistics that will be used later to construct statistical tests.

\subsection{Repeated measures ANOVA model for real data}
Let us recall the repeated measures ANOVA model and standard F-type test statistic. Suppose that there are $n$ subjects (e.g., patients), measured for $\ell\geq 2$ experimental conditions or time points (e.g., before, during, and after treatment). Assume that $Y_{ij}$ is an observation concerning the $i$-th object (repetition of an experiment) applied to the $j$-th subject, $i=1,\dots,\ell$, $j=1,\dots,n$. Let $N=n\ell$ be the number of all observations. In standard repeated measures ANOVA, the model $Y_{ij}=\mu+\alpha_i+\beta_j+e_{ij}$ is used, where $\mu$ is the general mean, $\alpha_i$ is the fixed effect of the $i$-th object, $\beta_j$ is a random effect of the $j$-th subject, and $e_{ij}$ is a random error. Of course, there are usual assumptions about the distributions of random elements of this model, but we do not specify them, as they will not be needed for our tests. The main aim of this model is to test the null hypothesis $\mathcal{H}_0:\alpha_1=\dots=\alpha_{\ell}=0$ asserting the non-significance of the objects, against $\mathcal{H}_1:\neg \mathcal{H}_0$. The F-type test statistic for these hypotheses is of the form:
\begin{equation}
\label{F_stat}
F=\frac{\mathrm{SSA}/(\ell-1)}{\mathrm{SSR}/((\ell-1)(n-1))},
\end{equation}
where $\mathrm{SSA}=n\sum_{i=1}^\ell(\bar{Y}_{i\cdot}-\bar{Y})^2$ is the sum of squares due to the hypothesis, $\mathrm{SSR}=\sum_{i=1}^\ell\sum_{j=1}^n(Y_{ij}-\bar{Y}_{i\cdot}-\bar{Y}_{\cdot j}+\bar{Y})^2$ is the sum of squares due to residuals or errors, $\bar{Y}_{i\cdot}=n^{-1}\sum_{j=1}^nY_{ij}$, $\bar{Y}_{\cdot j}=\ell^{-1}\sum_{i=1}^\ell Y_{ij}$, and $\bar{Y}=N^{-1}\sum_{i=1}^\ell\sum_{j=1}^nY_{ij}$. The null hypothesis is rejected for large values of $F$. The $F$ test statistic can be constructed in functional form and used for repeated measures analysis for functional data, as we describe in the next sections.

\subsection{Repeated measures ANOVA model for functional data}
Now, we describe the functional repeated measures analysis problem. It is a little different from that for real variables, which follows from the special characteristics of the functional data framework. Similarly as above, we have $n$ subjects subjected to $\ell\geq 2$ (possibly) different conditions, but the results of the experiments are functional observations. Namely, let the subjects be represented by a functional sample consisting of independent stochastic processes $Y_1,\dots,Y_n$ defined on the interval $[0,\ell]$. We assume that these data satisfy the following model proposed by Mart\'inez-Camblor and Corral (2011):
\begin{equation}
\label{Ymepsilon}
Y_j(t)=\mu(t)+e_j(t),\ j=1,\dots,n,\ t\in[0,\ell],
\end{equation}
where $\mu$ is a fixed mean function, and $e_j$ is a random process with zero mean function and covariance function $\gamma(s,t)$, $s,t\in[0,\ell]$. In this notation, $t\in[0,1]$ corresponds to the first experimental condition, $t\in[1,2]$ to the second, and so on. Thus, in this model we ignore the possible time periods between repetitions of the experiment, but this does not mean that they do not exist. In the model~\eqref{Ymepsilon}, it is interesting to test the equality of $\ell$ mean functions corresponding to experimental conditions; namely, the global null hypothesis is as follows:
\begin{equation}
\label{H0fda}
\left\{\begin{array}{l}
\mathcal{H}_0:\mu(t)=\mu(t+1)=\dots=\mu(t+(\ell-1))\ \ \forall t\in[0,1],\\
\mathcal{H}_1:\neg \mathcal{H}_0.\\
\end{array}\right.
\end{equation}
By rejecting the null hypothesis $\mathcal{H}_0$ we determine the presence of significant differences in the mean functions corresponding to the experimental conditions. However, we do not know which conditions are significantly different and which are not. To solve this problem, one needs to perform a post hoc analysis. This is of particular practical interest.

\subsection{Pointwise test statistics}
For the global null hypothesis~\eqref{H0fda}, the tests given by Mart\'inez-Camblor and Corral (2011) and Smaga (2019) used the pointwise sum of squares due to the hypothesis:
\begin{equation}
\label{ssa_p}
\mathrm{SSA}_{point}(t)=n\sum_{i=1}^\ell(\bar{Y}_{i\cdot}(t)-\bar{Y}(t))^2,\ t\in[0,1],
\end{equation}
where $$\bar{Y}_{i\cdot}(t)=n^{-1}\sum_{j=1}^nY_j(t+(i-1)),\ \bar{Y}(t)=N^{-1}\sum_{i=1}^\ell\sum_{j=1}^nY_j(t+(i-1)),$$
$i=1,\dots,\ell$. This pointwise test statistic takes into account ``between-group variability'' only. In contrast, the following pointwise counterpart of the F-type test statistic~\eqref{F_stat} for $\mathcal{H}_0$ in~\eqref{H0fda} also uses ``within-group variability'':
\begin{equation}
\label{Ft}
F_{point}(t)=\frac{\mathrm{SSA}_{point}(t)/(\ell-1)}{\mathrm{SSR}_{point}(t)/((\ell-1)(n-1))},\ t\in[0,1],
\end{equation}
where $$\mathrm{SSR}_{point}(t)=\sum_{i=1}^\ell\sum_{j=1}^n(Y_j(t+(i-1))-\bar{Y}_{i\cdot}(t)-\bar{Y}_{\cdot j}(t)+\bar{Y}(t))^2$$ is the pointwise adaptation of $\mathrm{SSR}$ in~\eqref{F_stat}, and
$$\bar{Y}_{\cdot j}(t)=\ell^{-1}\sum_{i=1}^\ell Y_j(t+(i-1)),\ j=1,\dots,n.$$
Using such additional information usually results in more powerful testing procedures. This was observed for standard ANOVA (Zhang et al., 2019; Smaga and Zhang, 2020) and for a paired two-sample test (Smaga, 2020) for functional data.

Using the pointwise F-type test statistic~\eqref{Ft}, we can construct the pointwise F-type test for~\eqref{H0fda} similarly to the pointwise F-type test proposed by Ramsay and Silverman (2005) for standard ANOVA for functional data. Such a test rejects the null hypothesis when it rejects the hypothesis for any fixed $t\in[0,1]$. Unfortunately, the pointwise F-type test may be time-consuming, since it must be performed for all $t\in[0,1]$. Moreover, this test does not guarantee that the null hypothesis in~\eqref{H0fda} is significant overall for a given significance level even when the pointwise F-type test is significant for all $t\in[0,1]$ at the same significance level (see, for example, Zhang, 2013). Thus, in the next section, we propose global tests for~\eqref{H0fda}, which overcome this difficulty.

\section{Statistical tests}
\label{sec_3}
In this section, we construct testing procedures based on the pointwise F-type test statistic~\eqref{Ft}. To approximate the distribution of test statistics, we use different bootstrap and permutation methods.

\subsection{Test statistics} 
To obtain global test statistics for $\mathcal{H}_0$ in~\eqref{H0fda}, one has to aggregate the pointwise statistics \eqref{ssa_p} and \eqref{Ft} presented above into one real variable. In functional data analysis, the natural idea is integration, since the common assumption is that the elements of a functional sample $Y_1,\dots,Y_n$ belong to the Hilbert space $L_2(T)$ of square integrable functions on the interval $T$. For instance, to verify the null hypothesis $\mathcal{H}_0$ in~\eqref{H0fda}, Mart\'inez-Camblor and Corral (2011) and Smaga (2019) integrated $\mathrm{SSA}_{point}$ as given in \eqref{ssa_p}, i.e., $$\mathcal{C}_n(\ell)=\int_0^1\mathrm{SSA}_{point}(t)dt.$$ 

Another less common idea in functional data analysis is to use a supremum. For example, Zhang et al. (2019) and Smaga and Zhang (2020) developed this idea in the case of standard ANOVA for functional data. In this paper, we use both of these ideas, but in contrast to $\mathcal{C}_n(\ell)$, we apply the appropriate operations to the pointwise F-type test statistic \eqref{Ft} instead of the pointwise sum of squares due to hypothesis~\eqref{ssa_p}. Namely, we propose the following test statistics: $$\mathcal{D}_n(\ell)=\int_0^1F_{point}(t)dt,\quad\mathcal{E}_n(\ell)=\sup\limits_{t\in[0,1]}F_{point}(t).$$

For the paired two-sample problem for functional data, these test statistics reduce to those considered by Smaga (2020). Thus, the results of the present paper are in fact extensions of the results obtained for the case $\ell=2$ by Smaga (2020) to repeated measures analysis for $\ell\geq 2$ groups. In the following section, we present the construction of the tests based on the test statistics $\mathcal{D}_n(\ell)$ and $\mathcal{E}_n(\ell)$.

\subsection{Permutation and bootstrap tests}
\label{sec_3_2}
In this subsection, we discuss the methods of approximation of the null distributions of the test statistics $\mathcal{D}_n(\ell)$ and $\mathcal{E}_n(\ell)$, which lead to new testing procedures for $\mathcal{H}_0$ as in~\eqref{H0fda}. For $\ell=2$, Smaga (2020) proposed several methods for this purpose. Unfortunately, they do not perform equally well, as was shown in a simulation study. Namely, the tests based on methods of approximation of the asymptotic null distributions of the test statistics (e.g., the parametric bootstrap approach) were usually too liberal for small and moderate values of the number $n$ of subjects. Therefore, we do not consider such methods in this paper, but concentrate on permutation and bootstrap approaches. In general, these methods do not need particular assumptions about the distribution of the data and a large number of observations to give a good approximation of the null distribution of test statistics. Of course, not all resampling methods work well, as we will see in the simulation study.

As the first method, we consider the permutation approach proposed by Mart\'inez-Camblor and Corral (2011) and Smaga (2020). Let $\mathcal{T}_n(\ell)$ denote one of the test statistics $\mathcal{D}_n(\ell)$ and $\mathcal{E}_n(\ell)$ (or possibly $\mathcal{C}_n(\ell)$) and let $B$ be a large number (e.g., $B=1000$). The steps of this procedure are as follows: 

\begin{enumerate}
\item Compute $\mathcal{T}_n(\ell)$ for the original data $Y_1(t),\dots,Y_n(t)$, $t\in[0,\ell]$.

\item Create a permutation sample $Y_1^b(t),\dots,Y_n^b(t)$, $t\in[0,\ell]$ from the given data in the following way: for each $j=1,\dots,n$ separately, randomly permute the observations $Y_j(t),Y_j(t+1),\dots,Y_j(t+(\ell-1))$, $t\in[0,1]$ corresponding to $\ell$ experimental conditions, and use them to form the permuted observation $Y_j^b(t)$.

\item Repeat step 2 $B$ times to obtain $B$ independent permutation samples.

\item For each permutation sample, compute the value of the test statistic $\mathcal{T}_n(\ell)$. Denote these by $\mathcal{T}_{n,b}(\ell)$, $b=1,\dots,B$.

\item The final $p$-value of the permutation test is defined by 
$B^{-1}\sum_{b=1}^BI(\mathcal{T}_{n,b}(\ell)>\mathcal{T}_n(\ell))$, where $I(A)$ stands for the usual indicator function on a set $A$.
\end{enumerate}
We refer to this permutation test as $\mathcal{P}_1$ for short. The following approaches follow the same steps, but use different methods of constructing new samples in step~2 and possibly modified forms of the test statistics in step~4.

In the second permutation approach, step 2 is performed as follows: We draw $Y_1^b(t),\dots,Y_n^b(t)$ for $t\in[0,1]$ randomly without replacement from the set $$\mathcal{A}=\{Y_1(t),\dots,Y_n(t),Y_1(t+1),\dots,Y_n(t+1),\dots,Y_1(t+(\ell-1)),\dots,Y_n(t+(\ell-1))\}$$ for $t\in[0,1]$ containing all $N$ observations; after that we draw $Y_1^b(t),\dots,Y_n^b(t)$ for $t\in[1,2]$ randomly without replacement from the remaining elements in $\mathcal{A}$, and so on. We refer to this procedure as the $\mathcal{P}_2$ method. It is similar to the procedure considered by Konietschke and Pauly (2014, method (I)).

Let us turn now to three bootstrap approaches. The first is the nonparametric bootstrap approach considered by Mart\'inez-Camblor and Corral (2011) and Smaga (2020). We refer to it as the $\mathcal{B}_1$ method for short. In this approach, the second step of the above procedure selects independent bootstrap samples $Y_1^b(t),\dots,Y_n^b(t)$, $t\in[0,\ell]$ drawn with replacement from the original sample $Y_1(t),\dots,Y_n(t)$, $t\in[0,\ell]$. Moreover, we need to modify step 4; namely, the pointwise sum of squares due to the hypothesis $\mathrm{SSA}_{point}$ for bootstrap samples is calculated as follows: $$\mathrm{SSA}_{point}^b(t)=n\sum_{i=1}^\ell(\bar{Y}_{i\cdot}^b(t)-\bar{Y}_{i\cdot}(t)-\bar{Y}^b(t)+\bar{Y}(t))^2,$$ where $t\in[0,1]$, and $\bar{Y}_{i\cdot}^b(t)$ and $\bar{Y}^b(t)$ are the appropriate sample means computed on the bootstrap sample.

In step 2 of the second nonparametric bootstrap approach, we first have to center the observations. Namely, let $Y_{1,c}(t)=Y_1(t)-\bar{Y}_{\bullet}(t),\dots,Y_{n,c}(t)=Y_n(t)-\bar{Y}_{\bullet}(t)$ for $t\in[0,\ell]$, where $\bar{Y}_{\bullet}(t)=n^{-1}\sum_{j=1}^nY_j(t)$ is the sample mean function for the original sample $Y_1,\dots,Y_n$. Second, $Y_1^b(t),\dots,Y_n^b(t)$ for $t\in[0,1]$ are randomly drawn with replacement from $Y_{1,c}(t),\dots,Y_{n,c}(t)$, $t\in[0,1]$; independently, $Y_1^b(t),\dots,Y_n^b(t)$ for $t\in[1,2]$ are randomly drawn with replacement from $Y_{1,c}(t),\dots,Y_{n,c}(t)$, $t\in[1,2]$, and so on. We refer to this method for short as the $\mathcal{B}_2$ method. It is an extension of the methods considered by Konietschke and Pauly (2014) and Smaga (2020).

The last method considered is the parametric bootstrap approach (the $\mathcal{B}_3$ test for short), which uses bootstrap samples that are constructed in a much different way than in the above methods. The idea of this approach is similar to the parametric bootstrap proposed by Konietschke et al. (2015). In step 2, we generate the bootstrap samples $Y_1^b(t),\dots,Y_n^b(t)$, $t\in[0,\ell]$ from the Gaussian process with zero mean function and covariance function equal to the sample covariance function
\begin{equation}
\label{cov_fun_est}
\hat{\gamma}(s,t)=\frac{1}{n-1}\sum_{j=1}^n(Y_j(s)-\bar{Y}_{\bullet}(s))(Y_j(t)-\bar{Y}_{\bullet}(t)),\ s,t\in[0,\ell],
\end{equation}
which is the unbiased estimator of the covariance function $\gamma(s,t)$ (Zhang, 2013, page 108). Such bootstrap samples satisfy the null hypothesis $\mathcal{H}_0$ in~\eqref{H0fda}, since the mean function is the same for each $Y_j^b$. Moreover, the estimator of the covariance function used mimics the given covariance structure of the original observations, and it does this quite well, as we will see in the simulation study. Note that this parametric bootstrap method is different from the parametric bootstrap procedure considered by Mart\'inez-Camblor and Corral (2011) and Smaga (2020), which was used to approximate the asymptotic distributions of the test statistics $\mathcal{C}_n(\ell)$, $\mathcal{D}_n(2)$, and $\mathcal{E}_n(2)$.

\subsection{Post hoc analysis}
\label{sec_3_3}
After rejecting the global null hypothesis \eqref{H0fda}, it is of practical interest to perform post hoc analysis. The aim is to determine which experimental conditions give significantly different mean functions of the dependent variable $Y$, and which do not result in such differences. More precisely, we would like to test the family of null hypotheses:
\begin{equation}
\label{H0_post_hoc}
\left\{\begin{array}{l}
\mathcal{H}_0^{rs}:\mu(t+(r-1))=\mu(t+(s-1))\ \forall t\in[0,1],\\
\mathcal{H}_1^{rs}:\mu(t+(r-1))\neq\mu(t+(s-1))\ \text{for some}\ t\in[0,1],\\
\end{array}\right.
\end{equation}
for $r,s=1,\dots,\ell$, $r\neq s$. These hypotheses are also named pairwise comparisons. To test the family of local hypotheses~\eqref{H0_post_hoc}, we propose the following simple procedure:
\begin{enumerate}
\item Test each of the hypotheses in~\eqref{H0_post_hoc} using the data for the $r$-th and $s$-th objects, i.e., $Y_1(t),\dots,Y_n(t)$ for $t\in[r-1,r]$ and $t\in[s-1,s]$ respectively, and the chosen test from Section~\ref{sec_3_2}. Let $p_{rs}$ denote the $p$-values obtained.
\item Make a final decision using the Bonferroni method, i.e., reject the null hypothesis $\mathcal{H}_0^{rs}$ if $p_{rs}\leq \alpha/m$, where $m=\#\{p_{rs}:r,s=1,\dots,\ell, r\neq s\}$ is the number of null hypotheses considered.
\end{enumerate}
Step 2 of the above procedure can also be conducted by comparing the corrected $p$-values $p_{rs}^{Bonf}=m\cdot p_{rs}$ with the significance level $\alpha$, i.e., we reject $\mathcal{H}_0^{rs}$ if $p_{rs}^{Bonf}\leq \alpha$.

One of the common measures of the quality of multiple testing procedures is the family-wise error rate (FWER), namely the probability of making at least one type I error in the family of hypotheses (Tukey, 1953). It is known that the Bonferroni method controls the FWER at level $\alpha$, i.e., $\text{FWER}\leq\alpha$, if the individual tests control the type I error at level $\alpha$.

\section{Simulation study}
\label{sec_4}
In this section, we investigate the finite sample properties of the considered testing procedures in a Monte Carlo simulation study. We consider the control of the type I error level and the power of fifteen tests: the $\mathcal{P}_1$, $\mathcal{P}_2$, $\mathcal{B}_1$, $\mathcal{B}_2$, and $\mathcal{B}_3$ procedures based on the test statistics $\mathcal{C}_n(\ell)$, $\mathcal{D}_n(\ell)$, and $\mathcal{E}_n(\ell)$. We include the tests based on $\mathcal{C}_n(\ell)$ to provide a fair comparison between our tests and those based on the test statistic proposed by Mart\'inez-Camblor and Corral (2011). 

In simulation experiments, we consider $\ell=3$ repeated samples with different distributions, sample sizes, and amounts of correlation. Simulation studies in the case $\ell=2$ were conducted by Smaga (2020). In this case, the $\mathcal{P}_1$ tests based on the test statistics $\mathcal{D}_n(2)$ and $\mathcal{E}_n(2)$ performed best. For $\ell>2$, the situation may change, as we will describe. We consider testing of the global null hypothesis~\eqref{H0fda} as well as the pairwise comparisons~\eqref{H0_post_hoc}.

\subsection{Simulation design}
The $\ell=3$ repeated samples were generated according to model~\eqref{Ymepsilon}. We consider the following six main models specifying the mean functions, which directly indicate the truth or falsity of the null hypothesis ($t\in[0,1]$):
\begin{description}
\item[M1:]$\mu(t)=\mu(t+1)=\mu(t+2)=(\sin(2\pi t^2))^5$,
\item[M2:]$\mu(t)=\mu(t+1)=(\sin(2\pi t^2))^5$ and $\mu(t+2)=(\sin(2\pi t^2))^7$,
\item[M3:]$\mu(t)=\mu(t+1)=(\sin(2\pi t^2))^5$ and $\mu(t+2)=(\sin(2\pi t^2))^3$,
\item[M4:]$\mu(t)=\mu(t+1)=\mu(t+2)=\sqrt{6t/\pi}\exp(-6t)$,
\item[M5:]$\mu(t)=\mu(t+1)=\sqrt{6t/\pi}\exp(-6t)$ and $\mu(t+2)=\sqrt{13t/(2\pi)}\exp(-13t/2)$,
\item[M6:]$\mu(t)=\mu(t+1)=\sqrt{6t/\pi}\exp(-6t)$ and $\mu(t+2)=\sqrt{11t/(2\pi)}\exp(-11t/2)$.
\end{description}
These models are very similar to models considered in the simulation study by Mart\'inez-Camblor and Corral (2011). In models M1 and M4 (respectively M2, M3, M5, and M6), the null hypothesis $\mathcal{H}_0$ in~\eqref{H0fda} is true (respectively false), and hence we investigate the type I error level control (respectively power). We also considered two settings for the distribution of the data, namely normal and lognormal. In a normal setting, the errors were generated as follows: $e_j(t)=\xi B_{j1}(t)$, $e_j(t+1)=\rho e_j(t)+\xi (1-\rho^2)^{1/2}B_{j2}(t)$ and  $e_j(t+2)=\rho e_j(t+1)+\xi (1-\rho^2)^{1/2}B_{j3}(t)$ for $t\in[0,1]$ and $\xi=0.5,0.05$ for models M1--M3 and M4--M6 respectively. Here $B_{j1}$, $B_{j2}$, and $B_{j3}$ are the independent standard Brownian bridges, $j=1,\dots,n$. We set $\rho=0,0.25,0.5,0.75$ for zero, small, moderate, and large correlation respectively. The lognormal errors were the (appropriately centered) exponentially transformed errors of the normal setting, i.e., $\exp(e_j(t))$, $j=1,\dots,n$, $t\in[0,3]$. The sample sizes were set to $n=35,50,75,100$. Since functional data are not usually continuously observed in practice, the trajectories of the observations $Y_1(t),\dots,Y_n(t)$, $t\in[0,3]$ were discretized at the design time points $t_1,\dots,t_{101},t_1+1,\dots,t_{101}+1,t_1+2,\dots,t_{101}+2$, where $t_k$, $k=1,\dots,101$ were equispaced in the interval $[0,1]$.

The significance level $\alpha$ was set to $5\%$. The empirical sizes and powers of the tests were estimated as the proportion of rejections of the null hypothesis based on $1000$ Monte Carlo runs. The $p$-values of the tests were calculated based on $B=1000$ bootstrap and permutation samples. The R environment (R Core Team, 2022) was used to conduct the simulation study and the real data analysis described in Section~\ref{sec_5}. The code is available from the corresponding author upon request.

\begin{table}[!t]
\caption[]{Empirical sizes (as percentages) of all tests obtained in model~M1.}
\label{table1}
\centering
%\begin{scriptsize}
\begin{tabular}{l rrrrr rrrrr rrrrr}
\hline\noalign{\smallskip}
& \multicolumn{5}{l}{$\mathcal{C}_n(3)$} &  \multicolumn{5}{l}{$\mathcal{D}_n(3)$} &   \multicolumn{5}{l}{$\mathcal{E}_n(3)$}\\%\cline{3-7}\cline{9-13}\cline{15-19}
$\rho$& $\mathcal{P}_1$&$\mathcal{P}_2$&$\mathcal{B}_1$&$\mathcal{B}_2$&$\mathcal{B}_3$  &  $\mathcal{P}_1$&$\mathcal{P}_2$&$\mathcal{B}_1$&$\mathcal{B}_2$&$\mathcal{B}_3$  &  $\mathcal{P}_1$&$\mathcal{P}_2$&$\mathcal{B}_1$&$\mathcal{B}_2$&$\mathcal{B}_3$\\
\hline\noalign{\smallskip}
%\noalign{\smallskip}\noalign{\smallskip}
&\multicolumn{2}{l}{normal}&&  &  &&&&  &  &&&&\\
&\multicolumn{2}{l}{$n=35$}&&  &  &&&&  &  &&&&\\
0.00&  4.9&5.4&5.0&5.5&4.9 & 5.1&5.5&3.6&4.4&4.5 & 4.1&4.2&2.9&3.4&3.3\\
0.25&  4.8&1.7&4.8&2.1&4.6 & 5.7&5.3&3.4&4.6&4.0 & 6.1&6.1&3.4&4.4&3.7\\
0.50&  5.3&0.5&4.8&0.8&4.5 & 5.2&4.5&3.7&4.7&4.0 & 6.7&7.1&3.7&6.0&3.9\\
0.75&  4.7&0.0&3.9&0.0&4.2 & 5.6&5.2&3.3&5.1&4.0 & 8.2&8.1&3.7&7.4&4.3\\
&\multicolumn{2}{l}{$n=50$}&&  &  &&&&  &  &&&&\\
0.00&  4.7&4.6&4.6&5.4&4.2 & 5.1&5.0&3.9&4.3&4.2 & 4.7&4.9&3.9&4.0&3.4\\
0.25&  5.3&2.8&5.3&2.6&4.8 & 5.7&5.7&4.7&5.2&4.9 & 5.6&5.9&4.5&5.4&4.4\\
0.50&  5.7&0.7&5.6&0.6&5.0 & 6.0&5.8&4.5&5.3&4.8 & 6.8&6.9&4.3&6.3&4.3\\
0.75&  5.8&0.0&5.3&0.0&5.0 & 5.8&5.5&4.8&5.2&5.0 & 9.4&9.1&5.3&8.9&5.2\\
&\multicolumn{2}{l}{$n=75$}&&  &  &&&&  &  &&&&\\
0.00& 5.0&5.2&4.5&5.3&5.0 & 4.6&5.1&4.3&4.7&4.6 & 4.7&5.1&3.9&4.2&3.4\\
0.25& 5.6&1.9&4.9&2.2&4.4 & 5.1&5.5&4.1&4.5&4.2 & 5.4&5.9&4.3&4.9&3.8\\
0.50& 5.6&0.4&5.1&0.4&5.0 & 5.7&5.3&4.8&5.1&4.8 & 6.5&6.4&4.3&6.6&3.9\\
0.75& 5.7&0.0&5.3&0.0&5.2 & 5.9&5.7&4.6&6.2&4.8 & 9.5&9.8&4.7&9.4&4.2\\
&\multicolumn{2}{l}{$n=100$}&&  &  &&&&  &  &&&&\\
0.00& 5.4&5.7&5.8&5.5&5.5 & 5.6&5.3&4.8&4.9&5.0 & 4.8&5.2&3.9&4.5&3.4\\
0.25& 5.3&2.2&5.1&2.2&5.2 & 5.5&5.4&5.0&5.5&4.9 & 5.6&5.6&4.0&4.9&3.4\\
0.50& 5.1&0.6&4.7&0.5&4.7 & 5.2&5.4&4.6&5.1&4.4 & 7.2&6.7&4.6&7.0&4.7\\
0.75& 6.1&0.0&5.3&0.0&5.7 & 5.5&5.8&4.7&5.3&5.1 & 9.4&9.4&4.1&9.5&4.1\\
&\multicolumn{2}{l}{lognormal}&&  &  &&&&  &  &&&&\\
&\multicolumn{2}{l}{$n=35$}&&  &  &&&&  &  &&&&\\
0.00& 4.4&4.8&3.4&4.4&3.5 & 4.4&4.7&3.0&3.7&3.5 & 5.4&6.0&3.3&4.3&4.1\\
0.25& 4.6&1.8&3.5&1.7&3.8 & 4.7&5.2&3.3&4.0&3.7 & 5.4&6.0&3.5&4.4&4.4\\
0.50& 5.0&0.1&3.6&0.1&3.4 & 5.4&5.3&3.0&4.4&3.5 & 7.2&7.1&3.7&5.7&4.6\\
0.75& 5.6&0.0&4.0&0.0&3.4 & 5.7&5.6&2.9&4.3&3.5 & 9.3&9.1&4.0&7.8&5.0\\
&\multicolumn{2}{l}{$n=50$}&&  &  &&&&  &  &&&&\\
0.00& 4.8&5.0&4.4&4.6&4.1 & 4.8&5.3&3.6&4.0&3.9 & 5.1&5.2&3.7&4.1&4.4\\
0.25& 6.0&2.1&4.3&1.9&4.8 & 5.4&5.5&3.8&4.4&4.3 & 5.9&5.9&4.1&5.0&4.5\\
0.50& 6.5&0.3&5.1&0.4&5.0 & 6.3&6.0&4.4&5.1&4.6 & 7.9&7.4&5.0&6.5&5.9\\
0.75& 5.8&0.0&4.2&0.0&4.0 & 5.6&5.7&3.2&5.4&3.9 & 9.6&9.0&4.2&8.6&4.9\\
&\multicolumn{2}{l}{$n=75$}&&  &  &&&&  &  &&&&\\
0.00& 5.4&5.9&4.9&5.0&4.4 & 5.3&5.3&4.2&4.4&4.2 & 5.4&5.5&4.3&4.7&4.5\\
0.25& 5.7&2.2&4.2&2.3&5.1 & 5.8&5.6&4.1&5.1&4.8 & 5.0&5.3&3.8&4.3&3.7\\
0.50& 6.2&0.4&5.3&0.4&5.3 & 6.3&6.5&4.8&5.5&5.0 & 8.2&7.8&4.7&7.3&5.4\\
0.75& 6.3&0.0&5.0&0.0&5.2 & 6.4&5.9&4.9&6.3&5.0 & 9.6&8.9&4.6&8.9&5.4\\
&\multicolumn{2}{l}{$n=100$}&&  &  &&&&  &  &&&&\\
0.00& 5.3&5.9&4.8&5.7&5.1 & 5.3&5.6&4.4&5.1&4.8 & 5.3&5.4&4.4&4.9&4.8\\
0.25& 5.2&1.7&4.8&1.6&5.1 & 5.6&5.7&4.7&5.2&5.0 & 5.9&6.2&4.6&5.8&5.1\\
0.50& 3.5&0.1&3.2&0.0&3.0 & 3.7&3.7&3.1&3.4&2.9 & 6.2&5.9&3.4&5.4&3.5\\
0.75& 4.5&0.0&4.1&0.0&3.9 & 4.5&4.3&3.9&4.3&3.7 & 7.5&7.9&3.2&7.5&3.6\\
\noalign{\smallskip}\hline\noalign{\smallskip}
\end{tabular}
%\end{scriptsize}
\end{table}

{\small\tabcolsep=6.7pt
\begin{table}[!t]
\caption[]{Empirical powers (as percentages) of all tests obtained in selected cases in models~M2, M3, M5, and M6.}% Empirical powers in the omitted rows for $n=75,100$ are always 100\%.}
\label{table2}
\centering
%\begin{footnotesize}
\begin{scriptsize}
\begin{tabular}{l rrrrr rrrrr rrrrr}%{p{0.3cm}p{0.11cm}p{0.62cm}p{0.62cm}p{0.62cm}p{0.62cm}p{0.62cm}p{0.02cm}p{0.62cm}p{0.62cm}p{0.62cm}p{0.62cm}p{0.62cm}p{0.02cm}p{0.62cm}p{0.62cm}p{0.62cm}p{0.62cm}p{0.62cm}}
\hline\noalign{\smallskip}
& \multicolumn{5}{l}{$\mathcal{C}_n(3)$}  & \multicolumn{5}{l}{$\mathcal{D}_n(3)$}  & \multicolumn{5}{l}{$\mathcal{E}_n(3)$}\\%\cline{2-6}\cline{7-11}\cline{12-16}
$\rho$&  $\mathcal{P}_1$&$\mathcal{P}_2$&$\mathcal{B}_1$&$\mathcal{B}_2$&$\mathcal{B}_3$  & $\mathcal{P}_1$&$\mathcal{P}_2$&$\mathcal{B}_1$&$\mathcal{B}_2$&$\mathcal{B}_3$  &  $\mathcal{P}_1$&$\mathcal{P}_2$&$\mathcal{B}_1$&$\mathcal{B}_2$&$\mathcal{B}_3$\\
\hline\noalign{\smallskip}
%\noalign{\smallskip}\noalign{\smallskip}
\multicolumn{6}{l}{Model 2 normal}&&&&  &  &&&&\\
&\multicolumn{2}{l}{$n=35$}&&  &  &&&&  &  &&&&\\
0.00& 32.5&32.1&30.6&32.0&30.0 & 45.0&45.1&34.8&38.4&38.5 & 88.7&88.3&84.4&85.8&84.5\\
0.25& 42.9&24.7&40.6&24.5&40.2 & 59.5&59.0&47.6&54.0&52.1 & 96.1&95.9&92.1&94.8&93.3\\
0.50& 65.6&13.7&59.3&14.8&59.6 & 82.1&81.9&72.3&78.8&74.0 & 99.9&100&98.9&99.7&99.4\\
0.75& 97.8&5.3&96.6&5.7&96.2 & 99.9&99.8&98.7&99.7&99.5 & 100&100&100&100&100\\
&\multicolumn{2}{l}{$n=50$}&&  &  &&&&  &  &&&&\\
0.00& 52.6&53.1&51.0&53.4&50.9 & 71.3&71.2&64.7&68.3&66.8 & 98.9&99.2&98.7&98.9&98.9\\
0.25& 69.9&43.0&65.2&45.0&65.5 & 85.3&85.5&78.8&82.4&80.7 & 99.9&99.9&99.9&99.9&99.8\\
0.50& 91.0&33.5&87.1&33.4&87.9 & 98.3&97.9&96.2&97.8&96.7 & 100&100&100&100&100\\
0.75& 100&16.4&99.9&18.2&99.9 & 100&100&100&100&100 & 100&100&100&100&100\\
&\multicolumn{2}{l}{$n=75$}&&  &  &&&&  &  &&&&\\
0.00& 82.8&82.2&80.6&82.7&80.5 & 94.8&95.5&94.0&94.8&94.6 & 100&100&100&100&100\\
0.25& 94.0&78.1&92.5&77.0&93.0 & 98.9&99.0&98.1&99.1&98.5 & 100&100&100&100&100\\
0.50& 99.5&70.6&99.5&71.2&99.7 & 100&100&100&100&100 & 100&100&100&100&100\\
0.75& 100&62.2&100&62.6&100 & 100&100&100&100&100 & 100&100&100&100&100\\
&\multicolumn{2}{l}{$n=100$}&&  &  &&&&  &  &&&&\\
0.00& 97.5&97.3&96.9&97.1&97.4 & 99.8&99.8&99.8&99.8&99.9 & 100&100&100&100&100\\
0.25& 99.8&96.5&99.6&96.8&99.6 & 100&100&100&100&100 & 100&100&100&100&100\\
0.50& 100&95.7&100&95.2&100 & 100&100&100&100&100 & 100&100&100&100&100\\
0.75& 100&95.2&100&94.6&100 & 100&100&100&100&100 & 100&100&100&100&100\\
\multicolumn{6}{l}{Model 2 lognormal}&&&&  &  &&&&\\
&\multicolumn{2}{l}{$n=35$}&&  &  &&&&  &  &&&&\\
0.00& 86.8&87.3&82.9&86.4&82.3 & 86.7&86.7&79.5&83.4&81.7 & 95.5&95.7&92.4&93.6&94.4\\
0.25& 94.5&84.1&91.3&82.5&92.0 & 94.2&94.1&89.2&91.8&91.4 & 98.7&98.8&97.7&98.6&98.5\\
0.50& 99.1&78.3&98.7&77.1&98.6 & 99.1&99.1&98.3&99.1&98.5 & 99.9&100&99.8&100&99.8\\
0.75& 100&68.7&100&69.6&100 & 100&100&100&100&100 & 100&100&100&100&100\\
&\multicolumn{2}{l}{$n=50$}&&  &  &&&&  &  &&&&\\
0.00& 98.6&98.6&97.7&98.5&97.8 & 98.4&98.4&97.1&97.7&97.5 & 99.7&99.6&99.6&99.6&99.7\\
0.25& 99.5&97.8&99.5&97.8&99.4 & 99.5&99.5&99.1&99.6&99.3 & 100&100&100&100&100\\
0.50& 100&97.3&100&97.3&100 & 100&100&100&100&100 & 100&100&100&100&100\\
0.75& 100&97.5&100&97.3&100 & 100&100&100&100&100 & 100&100&100&100&100\\
%&&\multicolumn{2}{l}{$n=75$}&&  &&  &&&&  &&  &&&&\\
%0.00&& 100&100&100&100&100 && 100&100&100&100&100 && 100&100&100&100&100\\
%0.25&& 100&100&100&100&100 && 100&100&100&100&100 && 100&100&100&100&100\\
%0.50&& 100&100&100&100&100 && 100&100&100&100&100 && 100&100&100&100&100\\
%0.75&& 100&100&100&100&100 && 100&100&100&100&100 && 100&100&100&100&100\\
%&&\multicolumn{2}{l}{$n=100$}&&  &&  &&&&  &&  &&&&\\
%0.00&& 100&100&100&100&100 && 100&100&100&100&100 && 100&100&100&100&100\\
%0.25&& 100&100&100&100&100 && 100&100&100&100&100 && 100&100&100&100&100\\
%0.50&& 100&100&100&100&100 && 100&100&100&100&100 && 100&100&100&100&100\\
%0.75&& 100&100&100&100&100 && 100&100&100&100&100 && 100&100&100&100&100\\
\multicolumn{6}{l}{Model 3 normal}&&&&  &  &&&&\\
&\multicolumn{2}{l}{$n=35$}&&  &  &&&&  &  &&&&\\
0.00 & 92.9 & 93.6 & 91.1 & 93.1 & 91.5 & 99.3 & 99.7 & 98.2 & 98.9 & 99.2
& 100 & 100 & 100 & 100 & 100 \\
0.25 & 98.3 & 91.3 & 97.6 & 91.2 & 97.5 & 99.8 & 99.8 & 99.4 & 99.8 &
99.7 & 100 & 100 & 100 & 100 & 100 \\
0.50 & 99.8 & 90.9 & 99.7 & 91.0 & 99.8 & 100 & 100 & 100 & 100 & 100 & 100
& 100 & 100 & 100 & 100 \\
0.75 & 100 & 88.4 & 100 & 89.3 & 100 & 100 & 100 & 100 & 100 & 100 & 100
& 100 & 100 & 100 & 100 \\
%&\multicolumn{2}{l}{$n=50$}&&  &  &&&&  &  &&&&\\
%0.00 & 99.9 & 99.9 & 99.6 & 99.9 & 99.9 & 100 & 100 & 100 & 100 & 100 & 100
%& 100 & 100 & 100 & 100 \\
%0.25 & 100 & 99.9 & 100 & 99.7 & 100 & 100 & 100 & 100 & 100 & 100 & 100
%& 100 & 100 & 100 & 100 \\
%0.50 & 100 & 99.9 & 100 & 100 & 100 & 100 & 100 & 100 & 100 & 100 & 100 &
%100 & 100 & 100 & 100 \\
%0.75 & 100 & 99.8 & 100 & 99.9 & 100 & 100 & 100 & 100 & 100 & 100 & 100
%& 100 & 100 & 100 & 100 \\
\multicolumn{6}{l}{Model 5 normal}&&&&  &  &&&&\\
&\multicolumn{2}{l}{$n=35$}&&  &  &&&&  &  &&&&\\
0.00 & 55.1 & 54.4 & 54.1 & 54.2 & 54.1 & 63.2 & 63.4 & 55.2 & 60.0 & 60.0 & 91.0
& 91.1 & 86.7 & 88.7 & 88.2 \\
0.25 & 61.3 & 47.5 & 61.8 & 49.2 & 61.8 & 72.4 & 72.2 & 66.2 & 70.6 &
69.1 & 96.6 & 96.8 & 94.1 & 96.2 & 93.8 \\
0.50 & 77.6 & 40.6 & 76.7 & 42.2 & 75.8 & 89.0 & 89.2 & 83.0 & 87.5 & 85.6 &
98.9 & 99.0 & 97.9 & 98.6 & 98.3 \\
0.75 & 97.3 & 34.3 & 97.0 & 37.8 & 97.1 & 99.8 & 99.8 & 99.4 & 99.8 & 99.6
& 100 & 100 & 100 & 100 & 100 \\
&\multicolumn{2}{l}{$n=50$}&&  &  &&&&  &  &&&&\\
0.00 & 72.5 & 72.6 & 73.3 & 73.0 & 73.6 & 81.4 & 82.2 & 78.8 & 80.2 & 80.1 &
98.9 & 99.0 & 98.4 & 98.6 & 98.3 \\
0.25 & 82.3 & 69.9 & 81.1 & 70.7 & 81.1 & 91.4 & 91.1 & 88.7 & 89.7 &
88.9 & 99.7 & 99.7 & 99.6 & 99.5 & 99.5 \\
0.50 & 92.4 & 66.9 & 92.2 & 69.0 & 92.0 & 97.5 & 97.3 & 96.7 & 97.2 & 97.0 &
100 & 100 & 100 & 100 & 100 \\
0.75 & 100 & 63.9 & 99.9 & 67.0 & 99.9 & 100 & 100 & 100 & 100 & 100 & 100
& 100 & 100 & 100 & 100 \\
%&\multicolumn{2}{l}{$n=75$}&&  &  &&&&  &  &&&&\\
%0.00 & 90.4 & 91.1 & 90.8 & 91.3 & 89.9 & 96.3 & 96.9 & 95.9 & 96.1 & 95.9
%& 100 & 100 & 100 & 100 & 100 \\
%0.25 & 95.5 & 91.2 & 95.7 & 91.4 & 95.4 & 99.3 & 99.2 & 99.1 & 99.2 &
%99.2 & 100 & 100 & 100 & 100 & 100 \\
%0.50 & 99.2 & 90.2 & 98.8 & 90.2 & 99.1 & 99.9 & 99.9 & 99.9 & 99.9 &
%99.9 & 100 & 100 & 100 & 100 & 100 \\
%0.75 & 100 & 93.8 & 100 & 93.9 & 100 & 100 & 100 & 100 & 100 & 100 & 100
%& 100 & 100 & 100 & 100 \\
%&\multicolumn{2}{l}{$n=100$}&&  &  &&&&  &  &&&&\\
%0.00 & 96.7 & 96.6 & 96.8 & 97.0 & 96.7 & 99.2 & 99.3 & 99.2 & 99.2 & 99.3 &
%100 & 100 & 100 & 100 & 100 \\
%0.25 & 99.0 & 97.0 & 99.0 & 97.0 & 99.0 & 99.9 & 100 & 99.9 & 99.9 & 99.9 & 100 &
%100 & 100 & 100 & 100 \\
%0.50 & 99.9 & 97.3 & 99.9 & 97.6 & 99.9 & 100 & 100 & 100 & 100 & 100 &
%100 & 100 & 100 & 100 & 100 \\
%0.75 & 100 & 99.6 & 100 & 99.8 & 100 & 100 & 100 & 100 & 100 & 100 & 100
%& 100 & 100 & 100 & 100 \\
\multicolumn{6}{l}{Model 6 normal}&&&&  &  &&&&\\
&\multicolumn{2}{l}{$n=35$}&&  &  &&&&  &  &&&&\\
0.00 & 67.5 & 68.4 & 67.1 & 68.3 & 67.5 & 75.4 & 75.9 & 70.7 & 72.9 & 72.6
& 96.7 & 96.4 & 93.7 & 95.1 & 94.5 \\
0.25 & 71.9 & 60.7 & 72.5 & 62.9 & 71.9 & 82.1 & 82.3 & 77.2 & 80.7 &
79.4 & 98.6 & 99.0 & 98.0 & 98.4 & 97.8 \\
0.50 & 87.5 & 57.9 & 86.2 & 60.7 & 86.5 & 95.2 & 94.8 & 92.0 & 94.0 & 93.2 &
100 & 99.9 & 99.6 & 99.9 & 99.9 \\
0.75 & 98.8 & 52.7 & 98.6 & 56.2 & 98.9 & 100 & 100 & 100 & 100 & 100 &
100 & 100 & 100 & 100 & 100 \\
&\multicolumn{2}{l}{$n=50$}&&  &  &&&&  &  &&&&\\
0.00 & 84.3 & 84.4 & 83.6 & 84.7 & 83.4 & 91.7 & 91.4 & 89.1 & 90.4 & 90.8
& 99.9 & 99.9 & 99.9 & 99.7 & 99.7 \\
0.25 & 89.1 & 81.6 & 88.2 & 82.2 & 88.1 & 95.3 & 95.0 & 94.1 & 94.8 & 94.7
& 99.9 & 99.9 & 99.9 & 99.9 & 99.9 \\
0.50 & 96.2 & 82.1 & 95.7 & 82.6 & 96.2 & 99.4 & 99.8 & 98.9 & 99.3 &
99.2 & 100 & 100 & 100 & 100 & 100 \\
0.75 & 100 & 83.3 & 100 & 85.3 & 100 & 100 & 100 & 100 & 100 & 100 & 100
& 100 & 100 & 100 & 100 \\
%&\multicolumn{2}{l}{$n=75$}&&  &  &&&&  &  &&&&\\
%0.00 & 95.7 & 95.8 & 95.6 & 96.1 & 95.8 & 99.6 & 99.3 & 99 & 99.2 & 99.2 &
%100 & 100 & 100 & 100 & 100 \\
%0.25 & 98.5 & 96.8 & 98.4 & 96.7 & 98.6 & 99.9 & 99.9 & 99.9 & 99.9 &
%99.9 & 100 & 100 & 100 & 100 & 100 \\
%0.50 & 99.9 & 96.7 & 99.6 & 96.4 & 99.7 & 100 & 100 & 100 & 100 & 100 &
%100 & 100 & 100 & 100 & 100 \\
%0.75 & 100 & 99.1 & 100 & 99.0 & 100 & 100 & 100 & 100 & 100 & 100 & 100 &
%100 & 100 & 100 & 100 \\
\noalign{\smallskip}\hline\noalign{\smallskip}
\end{tabular}
\end{scriptsize}
%\end{footnotesize}
\end{table}
}

\subsection{Simulation results}
All results of the simulation study are contained in the supplement to this paper, which is available from the corresponding author upon request. Here, to save space, we present only selected results, which are nevertheless representative. They are given in Tables~\ref{table1}--\ref{table5}.

Let us first discuss the results of the simulation experiments for the fifteen tests given in Section~\ref{sec_3_2} for testing the global null hypothesis~\eqref{H0fda}. The results are presented in Tables~\ref{table1}--\ref{table2}. The type I error control is studied in models M1 (Table~\ref{table1}) and M4. Since the results for model M4 are similar to those for model M1, they are available in the supplement. For both normal and lognormal distributions, the empirical sizes have similar behavior. Although the $\mathcal{P}_1$ test with $\mathcal{E}_n(2)$ performed very well for $\ell=2$, it is too liberal when the number of groups is greater than two, especially for larger correlation ($\rho=0.5,0.75$). The same problem is observed for two other tests based on $\mathcal{E}_n(3)$, namely $\mathcal{P}_2$ and $\mathcal{B}_2$. Moreover, this bad behavior seems to diminish very slowly as the number of observations~$n$ increases. This means that these three tests should be used with great care. On the other hand, for any amount of correlation, the $\mathcal{C}_n(3)$-based $\mathcal{P}_2$ and $\mathcal{B}_2$ tests have small empirical sizes, which means that these testing procedures are conservative. They become more conservative with increasing correlation. Of course, such tests control the type I error level, but may have a larger type II error rate, or in other words, a smaller power. The remaining tests exhibit much better performance, although the $\mathcal{B}_1$ and $\mathcal{B}_3$ tests based on the test statistics $\mathcal{D}_n(3)$ and $\mathcal{E}_n(3)$ may have a slightly conservative character. 

For completeness, we consider all tests in the investigation of power. However, let us note that the results for the $\mathcal{P}_1$, $\mathcal{P}_2$, and $\mathcal{B}_2$ tests based on $\mathcal{E}_n(3)$ may be better than for the other tests due to the former tests’ being too liberal, making such a comparison unfair. The empirical powers of selected cases are presented in Table~\ref{table2}. For the other scenarios, the results are similar or close to 100\%. Naturally, the power increases when we have more data, i.e., as $n$ increases. Moreover, for almost all tests, this also holds for increasing correlation between consecutive samples, i.e., as $\rho$ increases. However, for the $\mathcal{C}_n(3)$-based $\mathcal{P}_2$ and $\mathcal{B}_2$ testing procedures, we usually observe the reverse behavior, which can be explained by their extreme conservative character. Furthermore, these two tests are the least powerful among all of the procedures considered when $\rho>0$. The remaining three tests based on $\mathcal{C}_n(3)$ perform much better, but they are usually less powerful than the $\mathcal{D}_n(3)$ procedures. In most cases, the permutation method outperforms the bootstrap approach for the tests based on $\mathcal{D}_n(3)$. The same seems to be true for the $\mathcal{C}_n(3)$ tests. Finally, the most powerful methods are the $\mathcal{E}_n(3)$ tests. This last observation was made in almost all of our experiments, but let us note that this is not always necessarily the case, as we will show below for pairwise comparisons.

{\small\tabcolsep=5.7pt
\begin{table}[!t]
\caption[]{Empirical FWER (as percentages) of all post hoc tests obtained in model~M1.}
\label{table3}
\centering
%\begin{scriptsize}
\begin{tabular}{l rrrrr rrrrr rrrrr}%{p{0.3cm}p{0.11cm}p{0.62cm}p{0.62cm}p{0.62cm}p{0.62cm}p{0.62cm}p{0.02cm}p{0.62cm}p{0.62cm}p{0.62cm}p{0.62cm}p{0.62cm}p{0.02cm}p{0.62cm}p{0.62cm}p{0.62cm}p{0.62cm}p{0.62cm}}
\hline\noalign{\smallskip}
& \multicolumn{5}{l}{$\mathcal{C}_n(3)$}  & \multicolumn{5}{l}{$\mathcal{D}_n(3)$}  & \multicolumn{5}{l}{$\mathcal{E}_n(3)$}\\%\cline{2-6}\cline{7-11}\cline{12-16}
$\rho$&  $\mathcal{P}_1$&$\mathcal{P}_2$&$\mathcal{B}_1$&$\mathcal{B}_2$&$\mathcal{B}_3$  & $\mathcal{P}_1$&$\mathcal{P}_2$&$\mathcal{B}_1$&$\mathcal{B}_2$&$\mathcal{B}_3$  &  $\mathcal{P}_1$&$\mathcal{P}_2$&$\mathcal{B}_1$&$\mathcal{B}_2$&$\mathcal{B}_3$\\
\hline\noalign{\smallskip}
&\multicolumn{2}{l}{normal}&&  &  &&&&  &  &&&&\\
&\multicolumn{2}{l}{$n=35$}&&  &  &&&&  &  &&&&\\
0.00 & 4.9 & 4.7 & 5.9 & 5.3 & 5.5 & 4.5 & 4.5 & 4.0 & 3.9 & 4.6 & 5.3 & 5.4
& 3.5 & 4.6 & 5.6 \\
0.25 & 4.4 & 2.3 & 5.4 & 2.4 & 4.9 & 4.2 & 3.8 & 3.3 & 3.6 & 3.6 & 4.1 &
4.3 & 2.4 & 3.6 & 4.1 \\
0.50 & 4.0 & 0.5 & 4.9 & 0.8 & 4.9 & 3.8 & 3.7 & 3.3 & 3.7 & 3.9 & 3.8 & 3.8
& 2.5 & 3.7 & 3.7 \\
0.75 & 5.4 & 0.1 & 6.4 & 0.1 & 6.4 & 5.2 & 5.9 & 4.5 & 4.8 & 5.6 & 5.4 &
5.3 & 3.3 & 5.1 & 5.0 \\
&\multicolumn{2}{l}{$n=50$}&&  &  &&&&  &  &&&&\\
0.00 & 4.8 & 4.4 & 5.4 & 5.6 & 6.0 & 4.9 & 4.2 & 3.7 & 4.1 & 4.0 & 4.8 & 4.3 &
3.4 & 3.4 & 4.3 \\
0.25 & 5.0 & 2.3 & 5.4 & 2.7 & 5.6 & 5.1 & 4.8 & 4.0 & 4.4 & 5.1 & 5.8 & 5.7 &
4.0 & 5.3 & 5.6 \\
0.50 & 4.2 & 0.2 & 5.3 & 0.6 & 5.9 & 4.1 & 4.5 & 4.0 & 4.1 & 4.2 & 4.5 & 4.7
& 3.1 & 4.5 & 4.2 \\
0.75 & 5.3 & 0.0 & 6.1 & 0.0 & 6.2 & 5.4 & 4.4 & 4.4 & 4.6 & 5.2 & 4.3 & 4.3 &
3.1 & 3.8 & 3.8 \\
&\multicolumn{2}{l}{$n=75$}&&  &  &&&&  &  &&&&\\
0.00 & 4.7 & 3.8 & 4.3 & 3.6 & 4.2 & 4.1 & 3.8 & 3.5 & 3.6 & 3.9 & 4.1 &
3.8 & 3.5 & 3.7 & 4.2 \\
0.25 & 5.6 & 1.8 & 6.2 & 2.3 & 6.3 & 5.7 & 5.4 & 5.5 & 5.1 & 5.6 & 4.3 &
4.4 & 3.4 & 4.0 & 4.4 \\
0.50 & 5.2 & 0.4 & 5.6 & 0.7 & 6.1 & 5.2 & 4.8 & 5.1 & 5.3 & 5.4 & 4.8 &
5.2 & 4.8 & 4.7 & 5.2 \\
0.75 & 3.6 & 0.0 & 4.4 & 0.0 & 4.5 & 4.4 & 3.9 & 3.7 & 3.9 & 3.8 & 3.9 & 4.3 &
4.2 & 4.3 & 3.8 \\
&\multicolumn{2}{l}{$n=100$}&&  &  &&&&  &  &&&&\\
0.00 & 4.8 & 5.1 & 5.1 & 4.8 & 4.7 & 5.1 & 5.0 & 5.0 & 4.9 & 5.0 & 4.8 & 4.5 &
4.7 & 4.7 & 4.6 \\
0.25 & 6.2 & 2.9 & 6.6 & 3.4 & 6.3 & 6.1 & 6.7 & 6.1 & 6.3 & 6.2 & 4.9 &
5.2 & 5.1 & 4.8 & 4.9 \\
0.50 & 5.7 & 0.9 & 5.7 & 1.3 & 6.2 & 5.2 & 5.2 & 5.4 & 5.2 & 5.5 & 5.1 &
4.8 & 4.4 & 5.5 & 4.6 \\
0.75 & 4.6 & 0.0 & 5.8 & 0.0 & 5.6 & 4.8 & 4.1 & 4.8 & 4.7 & 4.5 & 5.0 & 5.2 &
4.4 & 5.1 & 5.3 \\
&\multicolumn{2}{l}{lognormal}&&  &  &&&&  &  &&&&\\
&\multicolumn{2}{l}{$n=35$}&&  &  &&&&  &  &&&&\\
0.00 & 5.2 & 5.0 & 5.4 & 4.9 & 5.1 & 5.6 & 5.6 & 4.1 & 4.7 & 4.8 & 5.0 & 5.4 &
3.2 & 3.5 & 5.2 \\
0.25 & 5.2 & 2.7 & 4.7 & 2.1 & 4.6 & 4.5 & 4.6 & 3.2 & 3.6 & 4.7 & 4.2 &
4.3 & 2.4 & 2.8 & 4.3 \\
0.50 & 5.1 & 0.2 & 4.7 & 0.4 & 4.3 & 4.9 & 4.6 & 3.4 & 3.6 & 3.9 & 5.2 &
5.2 & 2.7 & 3.9 & 4.7 \\
0.75 & 4.3 & 0.0 & 4.1 & 0.0 & 3.7 & 4.0 & 3.9 & 3.0 & 3.7 & 3.8 & 4.5 & 4.5 & 2.6
& 3.6 & 4.4 \\
&\multicolumn{2}{l}{$n=50$}&&  &  &&&&  &  &&&&\\
0.00 & 3.8 & 4.4 & 4.3 & 4.4 & 3.7 & 3.6 & 4.4 & 3.3 & 4.0 & 3.6 & 4.4 & 4.1
& 2.5 & 3.4 & 4.2 \\
0.25 & 4.1 & 2.3 & 4.9 & 1.9 & 4.6 & 4.1 & 4.9 & 3.3 & 3.8 & 4.1 & 4.7 &
4.7 & 3.2 & 3.4 & 4.5 \\
0.50 & 5.7 & 0.6 & 5.7 & 0.5 & 5.7 & 6.4 & 5.7 & 4.9 & 5.6 & 6.3 & 6.2 &
5.8 & 3.7 & 5.0 & 6.3 \\
0.75 & 4.1 & 0.0 & 4.7 & 0.0 & 3.9 & 4.7 & 4.6 & 4.3 & 3.8 & 4.2 & 5.0 & 4.7 &
3.1 & 4.2 & 4.6 \\
&\multicolumn{2}{l}{$n=75$}&&  &  &&&&  &  &&&&\\
0.00 & 4.2 & 4.2 & 4.4 & 4.1 & 4.2 & 4.0 & 4.2 & 3.8 & 3.5 & 4.0 & 4.9 & 5.1 &
3.8 & 4.2 & 4.7 \\
0.25 & 4.9 & 2.2 & 4.6 & 2.4 & 4.0 & 4.7 & 4.8 & 4.4 & 4.5 & 4.6 & 5.4 & 5.7
& 4.3 & 5.0 & 5.0 \\
0.50 & 5.2 & 0.4 & 4.7 & 0.3 & 5.1 & 4.9 & 4.6 & 4.0 & 4.5 & 4.8 & 4.7 & 4.9
& 4.0 & 4.4 & 4.4 \\
0.75 & 4.5 & 0.0 & 3.7 & 0.0 & 3.9 & 4.3 & 4.7 & 3.2 & 4.0 & 3.7 & 4.7 & 4.1 &
3.5 & 3.8 & 4.2 \\
&\multicolumn{2}{l}{$n=100$}&&  &  &&&&  &  &&&&\\
0.00 & 4.7 & 4.7 & 4.4 & 4.7 & 4.6 & 4.7 & 4.5 & 4.4 & 4.4 & 4.2 & 4.3 &
4.5 & 4.2 & 4.2 & 5.0 \\
0.25 & 5.3 & 2.7 & 5.3 & 2.5 & 5.4 & 5.2 & 5.3 & 4.8 & 5.0 & 5.2 & 4.9 & 4.6
& 4.0 & 4.4 & 4.8 \\
0.50 & 3.7 & 0.5 & 3.3 & 0.3 & 3.3 & 4.0 & 3.7 & 3.2 & 3.7 & 3.2 & 5.3 & 5.4
& 4.5 & 5.1 & 5.4 \\
0.75 & 4.3 & 0.0 & 4.6 & 0.0 & 4.6 & 4.5 & 4.7 & 3.8 & 4.6 & 4.6 & 6.1 & 5.9 &
4.9 & 5.3 & 5.3 \\
\noalign{\smallskip}\hline\noalign{\smallskip}
\end{tabular}
%\end{scriptsize}
\end{table}
}

{\small\tabcolsep=5.3pt
\begin{table}[!t]
\caption[]{Empirical powers (as percentages) of all post hoc tests obtained in selected cases in model~M2.}% Empirical powers in the omitted rows are always 100\% or close to it.}
\label{table4}
\centering
\begin{footnotesize}
\begin{tabular}{ll rrrrr rrrrr rrrrr}%{p{0.3cm}p{0.11cm}p{0.62cm}p{0.62cm}p{0.62cm}p{0.62cm}p{0.62cm}p{0.02cm}p{0.62cm}p{0.62cm}p{0.62cm}p{0.62cm}p{0.62cm}p{0.02cm}p{0.62cm}p{0.62cm}p{0.62cm}p{0.62cm}p{0.62cm}}
\hline\noalign{\smallskip}
&&\multicolumn{5}{l}{$\mathcal{C}_n(3)$}  & \multicolumn{5}{l}{$\mathcal{D}_n3)$}  & \multicolumn{5}{l}{$\mathcal{E}_n(3)$}\\%\cline{2-6}\cline{7-11}\cline{12-16}
$\rho$&Comp&$\mathcal{P}_1$&$\mathcal{P}_2$&$\mathcal{B}_1$&$\mathcal{B}_2$&$\mathcal{B}_3$&$\mathcal{P}_1$&$\mathcal{P}_2$&$\mathcal{B}_1$&$\mathcal{B}_2$&$\mathcal{B}_3$&$\mathcal{P}_1$&$\mathcal{P}_2$&$\mathcal{B}_1$&$\mathcal{B}_2$&$\mathcal{B}_3$\\
\hline\noalign{\smallskip}
%& $N$ & & & & & & & & & & & & & & & \\
%m1 nor & & & & & & & & & & & & & & & & \\
\multicolumn{6}{l}{normal}&&&&  &  &&&&\\
$n=35$ & & & & & & & & & & & & & & & & \\
0.00 & 1-3 & 12.8 & 12.2 & 12.7 & 13.1 & 12.6 & 17.1 & 16.6 & 12.1 & 14.1 &
15.2 & 62.9 & 62.3 & 55.9 & 59.8 & 63.8 \\
& 2-3 & 14.7 & 14.7 & 15.0 & 14.7 & 14.5 & 20.7 & 19.3 & 15.9 & 16.4 &
18.7 & 63.5 & 62.7 & 56.0 & 60.1 & 64.5 \\
0.25 & 1-3 & 16.2 & 11.9 & 16.3 & 12.1 & 16.5 & 22.8 & 20.9 & 16.0 & 17.5 &
18.9 & 65.5 & 66.7 & 59.5 & 63.2 & 67.6 \\
& 2-3 & 22.8 & 9.0 & 21.7 & 8.8 & 22.4 & 33.1 & 30.6 & 24.5 & 26.7 & 28.6
& 80.3 & 80.4 & 73.1 & 78.6 & 81.8 \\
0.50 & 1-3 & 23.9 & 9.1 & 23.8 & 10.0 & 23.5 & 35.2 & 32.6 & 27.3 & 29.3 & 30.7
& 81.5 & 81.9 & 73.9 & 80.1 & 82.5 \\
& 2-3 & 49.4 & 5.3 & 47.1 & 6.4 & 46.6 & 65.4 & 62.5 & 55.3 & 58.7 &
60.1 & 97.0 & 96.5 & 94.9 & 96.5 & 97.1 \\
0.75 & 1-3 & 57.4 & 3.6 & 55.6 & 4.2 & 55.6 & 74.3 & 72.8 & 66.1 & 68.5 &
69.4 & 98.4 & 98.5 & 96.9 & 98.5 & 98.6 \\
& 2-3 & 94.3 & 1.0 & 93.0 & 1.6 & 92.8 & 99.1 & 98.7 & 97.5 & 98.4 & 98.0 &
100 & 100 & 100 & 100 & 100 \\
$n=50$ & & & & & & & & & & & & & & & & \\
0.00 & 1-3 & 26.7 & 25.9 & 26.9 & 25.9 & 26.1 & 39.1 & 36.0 & 32.7 & 33.8 & 35.0
& 90.3 & 90.1 & 87.5 & 89.3 & 90.8 \\
& 2-3 & 24.6 & 23.8 & 24.5 & 23.4 & 23.9 & 38.4 & 35.9 & 32.7 & 33.1 &
34.3 & 89.2 & 89.0 & 87.7 & 88.9 & 89.9 \\
0.25 & 1-3 & 25.3 & 19.9 & 25.6 & 21.1 & 25.5 & 39.7 & 38.7 & 31.4 & 35.2 &
36.8 & 91.8 & 91.1 & 89.8 & 91.4 & 92.0 \\
& 2-3 & 41.8 & 18.1 & 40.4 & 17.5 & 40.7 & 57.9 & 56.5 & 52.8 & 54.3 &
56.3 & 98.0 & 98.1 & 96.9 & 97.9 & 98.1 \\
0.50 & 1-3 & 41.1 & 18.3 & 40.7 & 18.8 & 41.5 & 59.7 & 57.5 & 53.3 & 55.1 &
56.9 & 96.8 & 96.6 & 95.9 & 96.4 & 97.0 \\
& 2-3 & 76.3 & 9.7 & 74.4 & 10.3 & 74.7 & 91.5 & 90.8 & 86.5 & 89.2 & 89.0
& 99.9 & 99.9 & 99.9 & 99.9 & 99.9 \\
0.75 & 1-3 & 86.0 & 8.9 & 84.1 & 9.4 & 85.0 & 96.9 & 96.7 & 94.8 & 96.5 & 95.6 &
100 & 100 & 100 & 100 & 100 \\
& 2-3 & 100 & 4.5 & 99.9 & 5.0 & 99.8 & 100 & 100 & 100 & 100 & 100 & 100
& 100 & 100 & 100 & 100 \\
$n=75$ & & & & & & & & & & & & & & & & \\
0.00 & 1-3 & 49.3 & 47.5 & 48.7 & 47.6 & 48.4 & 73.1 & 70.8 & 68.9 & 69.1 &
70.6 & 99.3 & 99.1 & 99.4 & 99.0 & 99.3 \\
& 2-3 & 50.4 & 49.6 & 48.1 & 48.4 & 48.5 & 72.9 & 72.2 & 68.5 & 68.1 &
67.8 & 99.4 & 99.4 & 99.4 & 99.5 & 99.5 \\
0.25 & 1-3 & 55.1 & 48.2 & 54.5 & 48.3 & 54.2 & 77.4 & 77.1 & 73.2 & 75.2 &
75.7 & 99.8 & 99.7 & 99.6 & 99.6 & 99.8 \\
& 2-3 & 75.1 & 44.2 & 74.2 & 43.5 & 73.4 & 92.7 & 91.9 & 89.5 & 90.6 &
89.9 & 100 & 100 & 100 & 100 & 100 \\
0.50 & 1-3 & 78.6 & 43.4 & 77.1 & 45.0 & 78.5 & 93.4 & 94.1 & 91.8 & 92.7 & 93.0
& 99.9 & 99.9 & 99.9 & 99.9 & 99.9 \\
& 2-3 & 98.1 & 37.4 & 97.5 & 38.4 & 97.7 & 99.6 & 99.6 & 99.5 & 99.6 &
99.6 & 100 & 100 & 100 & 100 & 100 \\
0.75 & 1-3 & 99.4 & 36.4 & 99.5 & 36.5 & 99.4 & 100 & 99.9 & 99.9 & 99.9 &
99.9 & 100 & 100 & 100 & 100 & 100 \\
& 2-3 & 100 & 31.5 & 100 & 33.0 & 100 & 100 & 100 & 100 & 100 & 100 & 100
& 100 & 100 & 100 & 100 \\
$n=100$ & & & & & & & & & & & & & & & & \\
0.00 & 1-3 & 75.9 & 78.0 & 75.8 & 77.2 & 75.4 & 92.9 & 93.5 & 92.2 & 92.0 & 92.1
& 100 & 100 & 100 & 100 & 100 \\
& 2-3 & 77.1 & 76.9 & 75.3 & 75.9 & 74.6 & 93.8 & 93.1 & 92.1 & 91.4 &
92.6 & 100 & 100 & 100 & 100 & 100 \\
0.25 & 1-3 & 80.9 & 74.9 & 79.8 & 75.2 & 79.4 & 95.5 & 95.6 & 94.0 & 94.6 &
94.5 & 100 & 100 & 100 & 100 & 100 \\
& 2-3 & 94.5 & 73.8 & 94.2 & 73.8 & 94.4 & 99.5 & 99.7 & 99.6 & 99.6 &
99.7 & 100 & 100 & 100 & 100 & 100 \\
0.50 & 1-3 & 96.3 & 76.6 & 96.1 & 77.0 & 95.7 & 100 & 99.9 & 99.9 & 99.7 & 100
& 100 & 100 & 100 & 100 & 100 \\
& 2-3 & 100 & 76.9 & 100 & 74.6 & 100 & 100 & 100 & 100 & 100 & 100 &
100 & 100 & 100 & 100 & 100 \\
0.75 & 1-3 & 100 & 75.4 & 100 & 74.2 & 100 & 100 & 100 & 100 & 100 & 100 &
100 & 100 & 100 & 100 & 100 \\
& 2-3 & 100 & 73.9 & 100 & 74.8 & 100 & 100 & 100 & 100 & 100 & 100 &
100 & 100 & 100 & 100 & 100 \\
%m1 log & & & & & & & & & & & & & & & & \\
\multicolumn{6}{l}{lognormal}&&&&  &  &&&&\\
$n=35$ & & & & & & & & & & & & & & & & \\
0.00 & 1-3 & 64.5 & 66.7 & 61.0 & 64.5 & 60.7 & 63.2 & 62.8 & 53.7 & 57.3 & 59.0
& 78.6 & 79.2 & 69.0 & 73.6 & 78.7 \\
& 2-3 & 65.6 & 64.5 & 62.1 & 63.2 & 62.0 & 64.1 & 63.1 & 55.3 & 58.1 &
59.3 & 80.1 & 80.0 & 70.2 & 74.7 & 79.8 \\
0.25 & 1-3 & 68.5 & 63.5 & 63.8 & 60.1 & 63.1 & 67.0 & 66.8 & 56.6 & 59.4 &
61.2 & 82.9 & 82.5 & 72.5 & 77.9 & 83.8 \\
& 2-3 & 83.7 & 62.1 & 79.8 & 61.0 & 79.6 & 82.9 & 82.4 & 75.4 & 77.9 &
79.2 & 92.2 & 92.4 & 86.2 & 89.7 & 92.3 \\
0.50 & 1-3 & 83.3 & 61.9 & 80.5 & 60.6 & 80.2 & 82.4 & 80.7 & 75.2 & 78.2 &
79.0 & 91.4 & 91.4 & 85.5 & 89.1 & 91.6 \\
& 2-3 & 98.1 & 58.6 & 97.6 & 57.1 & 97.5 & 97.8 & 98.3 & 95.5 & 97.0 &
97.2 & 99.3 & 99.2 & 98.6 & 99.1 & 99.2 \\
0.75 & 1-3 & 99.4 & 58.1 & 98.7 & 55.3 & 98.8 & 99.5 & 98.9 & 97.9 & 98.9 &
98.9 & 99.6 & 99.7 & 99.1 & 99.4 & 99.7 \\
& 2-3 & 100 & 50.9 & 100 & 50.4 & 100 & 100 & 100 & 100 & 100 & 100 &
100 & 100 & 100 & 100 & 100 \\
$n=50$ & & & & & & & & & & & & & & & & \\
0.00 & 1-3 & 88.8 & 90.2 & 87.5 & 88.2 & 87.9 & 88.8 & 89.1 & 85.9 & 86.7 &
87.3 & 95.6 & 95.7 & 94.3 & 95.0 & 95.4 \\
& 2-3 & 88.2 & 89.5 & 87.4 & 88.3 & 87.7 & 88.5 & 88.0 & 85.0 & 85.3 & 86.3
& 95.6 & 95.8 & 94.8 & 94.9 & 95.7 \\
0.25 & 1-3 & 92.2 & 88.8 & 90.8 & 88.1 & 90.3 & 92.4 & 91.3 & 88.4 & 88.7 &
90.1 & 97.1 & 96.6 & 95.5 & 96.3 & 96.9 \\
& 2-3 & 97.7 & 89.0 & 96.9 & 87.9 & 97.1 & 97.2 & 97.0 & 95.4 & 96.2 & 96.0 &
99.0 & 99.1 & 98.3 & 98.8 & 98.7 \\
0.50 & 1-3 & 97.5 & 88.6 & 96.4 & 86.9 & 96.7 & 97.8 & 97.7 & 95.3 & 96.5 &
96.5 & 99.4 & 99.6 & 99.0 & 99.4 & 99.4 \\
& 2-3 & 99.9 & 87.6 & 99.9 & 87.6 & 99.9 & 99.9 & 100 & 100 & 99.9 &
99.9 & 100 & 100 & 100 & 100 & 100 \\
0.75 & 1-3 & 100 & 88.5 & 100 & 87.4 & 100 & 100 & 100 & 100 & 100 & 100 &
100 & 100 & 100 & 100 & 100 \\
& 2-3 & 100 & 90.1 & 100 & 89.8 & 100 & 100 & 100 & 100 & 100 & 100 &
100 & 100 & 100 & 100 & 100 \\
\noalign{\smallskip}\hline\noalign{\smallskip}
\end{tabular}
\end{footnotesize}
\end{table}
}

{\small\tabcolsep=3.7pt
\begin{table}[!t]
\caption[]{Empirical powers (as percentages) of all post hoc tests obtained in selected cases in model~M5 under lognormal distribution.}% Empirical powers in the omitted rows are always 100\% or close to it.}
\label{table5}
%\begin{scriptsize}
\centering
\begin{tabular}{ll rrrrr rrrrr rrrrr}%{p{0.3cm}p{0.11cm}p{0.62cm}p{0.62cm}p{0.62cm}p{0.62cm}p{0.62cm}p{0.02cm}p{0.62cm}p{0.62cm}p{0.62cm}p{0.62cm}p{0.62cm}p{0.02cm}p{0.62cm}p{0.62cm}p{0.62cm}p{0.62cm}p{0.62cm}}
\hline\noalign{\smallskip}
&&\multicolumn{5}{l}{$\mathcal{C}_n(3)$}  & \multicolumn{5}{l}{$\mathcal{D}_n(3)$}  & \multicolumn{5}{l}{$\mathcal{E}_n(3)$}\\%\cline{2-6}\cline{7-11}\cline{12-16}
$\rho$&Comp &$\mathcal{P}_1$&$\mathcal{P}_2$&$\mathcal{B}_1$&$\mathcal{B}_2$&$\mathcal{B}_3$  & $\mathcal{P}_1$&$\mathcal{P}_2$&$\mathcal{B}_1$&$\mathcal{B}_2$&$\mathcal{B}_3$  &  $\mathcal{P}_1$&$\mathcal{P}_2$&$\mathcal{B}_1$&$\mathcal{B}_2$&$\mathcal{B}_3$\\
\hline\noalign{\smallskip}
%& $N$ & & & & & & & & & & & & & & & \\
$n=35$ & & & & & & & & & & & & & & & & \\
0.00 & 1-3 & 100 & 100 & 99.8 & 100 & 99.7 & 99.9 & 99.8 & 98.4 & 99.4 & 99.6
& 89.9 & 90.0 & 84.1 & 87.7 & 91.3 \\
& 2-3 & 99.9 & 100 & 99.7 & 99.8 & 99.7 & 99.8 & 99.8 & 98.9 & 98.8 &
99.2 & 89.4 & 89.5 & 82.9 & 86.6 & 89.6 \\
0.25 & 1-3 & 99.9 & 100 & 99.8 & 99.9 & 99.9 & 99.9 & 99.9 & 99.4 & 99.7 &
99.6 & 92.6 & 92.4 & 87.8 & 91.0 & 93.5 \\
& 2-3 & 100 & 100 & 100 & 99.9 & 100 & 100 & 100 & 100 & 100 & 100 &
98.2 & 97.7 & 96.7 & 97.4 & 97.9 \\
0.50 & 1-3 & 100 & 99.9 & 100 & 99.8 & 100 & 100 & 100 & 100 & 100 & 100 &
97.9 & 98.1 & 95.8 & 97.6 & 98.6 \\
& 2-3 & 100 & 100 & 100 & 100 & 100 & 100 & 100 & 100 & 100 & 100 & 99.9
& 99.9 & 99.8 & 99.8 & 99.9 \\
0.75 & 1-3 & 100 & 100 & 100 & 99.9 & 100 & 100 & 100 & 100 & 100 & 100 & 100
& 100 & 99.9 & 100 & 100 \\
& 2-3 & 100 & 100 & 100 & 100 & 100 & 100 & 100 & 100 & 100 & 100 & 100
& 100 & 100 & 100 & 100 \\
\noalign{\smallskip}\hline\noalign{\smallskip}
\end{tabular}
%\end{scriptsize}
\end{table}
}

Let us now turn to verifying the pairwise comparisons~\eqref{H0_post_hoc} (Tables~\ref{table3}--\ref{table5}). For the cases where all null hypotheses are true (models M1 and M4), we estimated the FWER (see Section~\ref{sec_3_3}). It should be controlled at level $\alpha$, i.e., $\text{FWER}\leq\alpha$. The results for model M1 are given in Table~\ref{table3}; those for model M4 are again similar, and hence are available in the supplement. Since we use just two groups in each comparison and the Bonferroni correction is applied, the FWER is controlled at the significance level by all tests considered. However, some of them again have a conservative character. This is evident for the $\mathcal{C}_n(3)$-based $\mathcal{P}_2$ and $\mathcal{B}_2$ testing procedures, but the $\mathcal{D}_n(3)$-based $\mathcal{P}_2$, $\mathcal{B}_1$, and $\mathcal{B}_2$ tests and $\mathcal{E}_n(3)$-based $\mathcal{B}_1$ and $\mathcal{B}_2$ tests are also slightly conservative. This results in loss of power (Table~\ref{table4}), especially for the $\mathcal{P}_2$ and $\mathcal{B}_2$ $\mathcal{C}_n(3)$ tests, except in the independent case ($\rho=0$). The empirical powers are presented for two comparisons, between the first and third samples and the second and third samples (denoted by 1-3 and 2-3 respectively), since in these cases the mean functions are different. The empirical powers show the good ability of the post hoc tests to detect differences in the mean functions for particular pairs of samples. In model M2 (Table~\ref{table4}) as well as models M3, M5, and M6 (data available in the supplement), we observe that the $\mathcal{E}_n(3)$ tests usually outperform the $\mathcal{D}_n(3)$ procedures, which are more powerful than the $\mathcal{C}_n(3)$ tests in most cases. However, we note that the opposite situation can sometimes apply. This is shown in Table~\ref{table5}, where the $\mathcal{E}_n(3)$ tests are less powerful than the remaining tests in a few cases. Except for the extremely conservative tests, the results for the 1-3 comparison are usually worse than those for the 2-3 comparison. This can be explained by the different amounts of correlation: the observations in samples 1 and 3 are less correlated than in samples 2 and 3, and as we observed in the investigation of powers for the global null hypothesis, greater correlation leads to larger power. The latter observation is also usually true for pairwise comparisons. 

To sum up, the best type I error control and power are achieved by the $\mathcal{E}_n(3)$-based $\mathcal{B}_1$ and $\mathcal{B}_3$ tests. However, the tests based on the test statistic $\mathcal{D}_n(3)$ (especially permutation tests) also have good finite sample properties.

\section{Real data application}
\label{sec_5}
In this section, we present an illustrative real data example and a simulation study based on a diffusion tensor imaging (DTI) data set. This example shows the practical application of the tests considered.

\subsection{Analysis of DTI data set}
\label{sec_5_1}
The DTI data were collected at Johns Hopkins University and the Kennedy Krieger Institute, and they are available in the R package \texttt{refund} (Goldsmith et al., 2022). The DTI is a magnetic resonance imaging technique providing different measures of water diffusivity along brain white matter tracts. It is used especially in diseases affecting the brain white matter tissue, such as multiple sclerosis (MS). In the experiment, DTI brain scans were recorded for MS patients at several visits (from 2 to 8). The aim was to assess the effect of neurodegeneration on disability. The DTI was used to determine the fractional anisotropy (FA) tract profiles for the corpus callosum (CCA).

\begin{figure}[t]
\includegraphics[width=0.99\textwidth,
height=0.2\textheight]{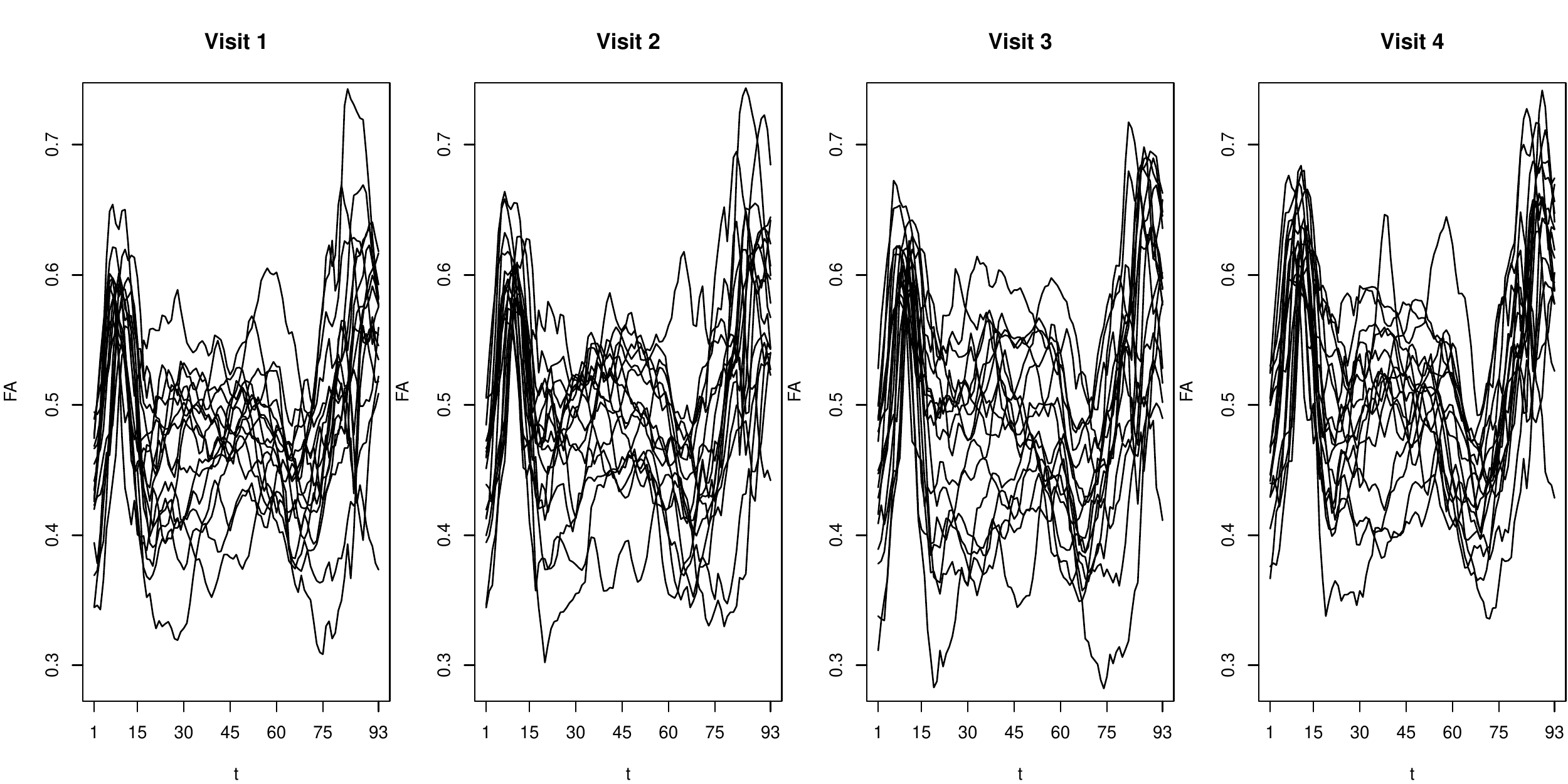}
\includegraphics[width=0.99\textwidth,
height=0.2\textheight]{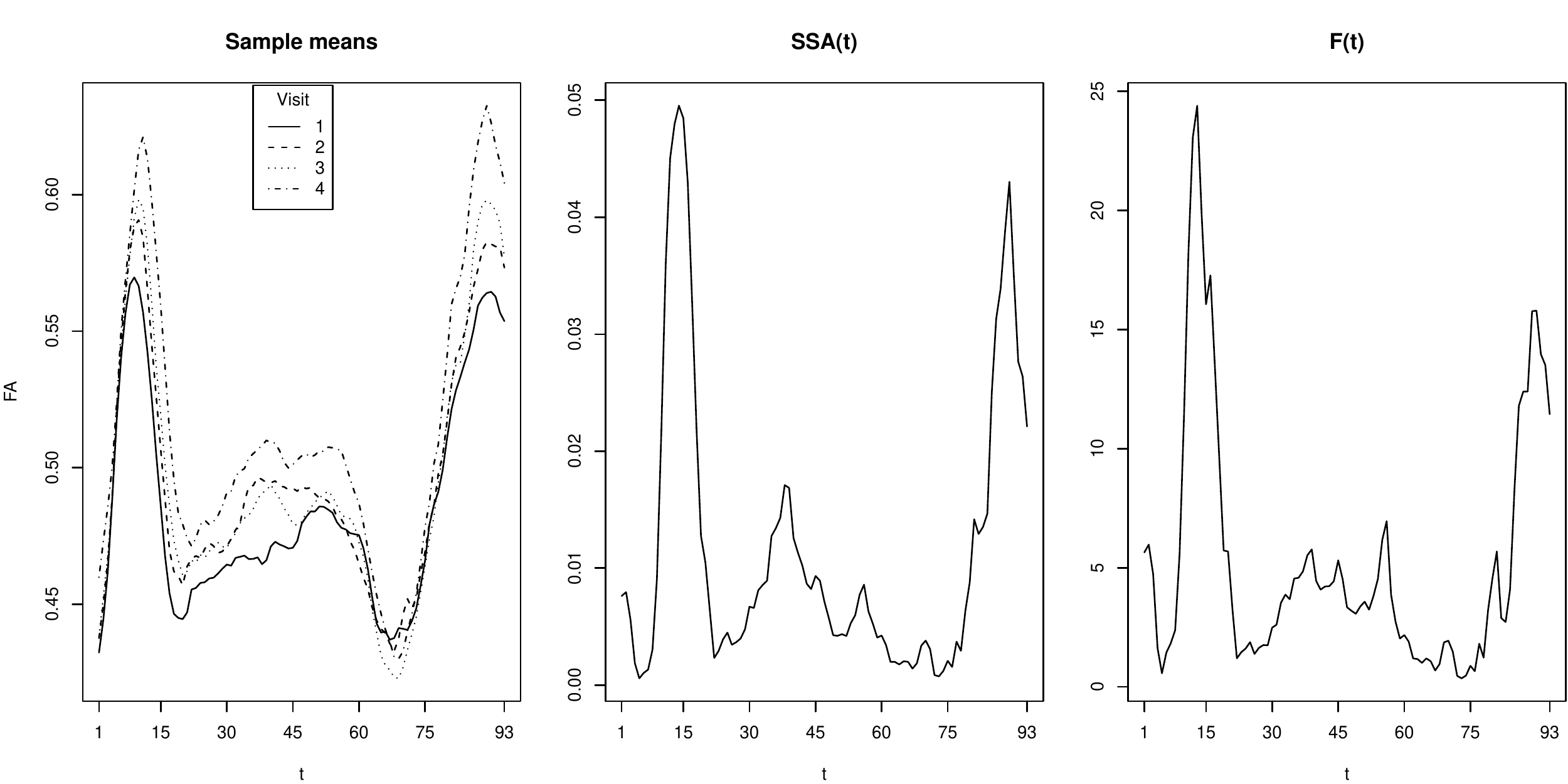}
\caption{Trajectories for the FA profiles at four visits (top panels); sample mean functions of the FA at different visits and pointwise test statistics (bottom panels).}
\label{Fig1}
\end{figure}

We first delete incomplete data in the DTI data set. For illustrative purposes, we then select the observations for $n=17$ patients (subjects) with $\ell=4$ successive visits (objects). Since the trajectories for the FA profiles are given at 93 design time points for each patient and each visit, we treat these data as functional data on the interval $[1,93]$. For each visit, they are presented in the top panels of Figure~\ref{Fig1}. It is of interest to check whether the FA profiles change significantly at different visits. Thus, we would like to test the equality of mean functions of the FA at different visits. First, we test the global null hypothesis~\eqref{H0fda}. The sample mean functions as well as the pointwise test statistics $\mathrm{SSA}_{point}(t)$ and $F_{point}(t)$ given in~\eqref{ssa_p} and~\eqref{Ft} respectively are depicted in the bottom panels of Figure~\ref{Fig1}. We observe that larger (respectively lower) values of these statistics correspond to larger (respectively smaller) differences between the sample mean functions. For the global null hypothesis~\eqref{H0fda}, the $p$-values of the $\mathcal{P}_2$ and $\mathcal{B}_2$ $\mathcal{C}_{17}(4)$ tests and the $\mathcal{B}_1$ $\mathcal{D}_{17}(4)$ test are 0.346, 0.285 and 0.001 respectively, while the $p$-values of the remaining tests considered in Section~\ref{sec_3} are equal to zero. The inadequate $p$-values of the $\mathcal{P}_2$ and $\mathcal{B}_2$ $\mathcal{C}_{17}(4)$ tests can be explained by the fact that these tests may be extremely conservative, which was observed in the simulation study based on the DTI data set presented in the next section (see also Section~\ref{sec_4}). The other tests reject the null hypothesis, indicating significant differences in the mean functions of the FA at different visits. From Figure~\ref{Fig1}, it seems that the values of these mean functions increase at subsequent visits. Since most of the tests rejected the global null hypothesis~\eqref{H0fda}, we know that there are significant differences between visits, but we do not know where exactly they occur. For this purpose, we perform tests for pairwise comparisons. We use Tukey's contrasts (Tukey, 1953), i.e., 1 vs. 2, 1 vs. 3, 1 vs. 4, 2 vs. 3, 2 vs. 4, and 3 vs. 4, where 1, 2, 3, and 4 are the numbers of the visits. Based on the procedure presented in Section~\ref{sec_3_3}, for each pair of visits, we apply the tests presented in Section~\ref{sec_3_2}, and correct the obtained $p$-values using the Bonferroni method. The corrected $p$-values are given in Table~\ref{table_rde_1}. We can easily observe that most of the tests (except the conservative ones) detect significant differences between visits 1 and 4 and between visits 2 and 4. Thus, it seems that the FA profiles for the last visit have much greater values than those for the first two visits. For the other comparisons, there are almost no significant differences. The rejections of equality of mean functions for 1 vs. 3 and 3 vs. 4 follow from the larger power of the corresponding tests (see the next section). The above observations seem to be in line with the expectation that the disease progresses over time.

{\small\tabcolsep=5pt
\begin{table}[t]
\caption[]{P-values (as percentages) of all tests for pairwise comparisons for the DTI data set.}
\label{table_rde_1}
%\begin{scriptsize}
\centering
\begin{tabular}{l rrrrr rrrrr rrrrr}
\hline\noalign{\smallskip}
&\multicolumn{5}{l}{$\mathcal{C}_{17}(4)$}  & \multicolumn{5}{l}{$\mathcal{D}_{17}(4)$}  & \multicolumn{5}{l}{$\mathcal{E}_{17}(4)$}\\%\cline{2-6}\cline{7-11}\cline{12-16}
Comp& $\mathcal{P}_1$&$\mathcal{P}_2$&$\mathcal{B}_1$&$\mathcal{B}_2$&$\mathcal{B}_3$  & $\mathcal{P}_1$&$\mathcal{P}_2$&$\mathcal{B}_1$&$\mathcal{B}_2$&$\mathcal{B}_3$  &  $\mathcal{P}_1$&$\mathcal{P}_2$&$\mathcal{B}_1$&$\mathcal{B}_2$&$\mathcal{B}_3$\\
\hline\noalign{\smallskip}
1-2&6.0&100&6.6&100&5.4&8.4&29.4&19.2&24.0&10.8&20.4&15.6&43.8&9.6&13.8\\
1-3&30.0&100&34.8&100&33.0&32.4&31.2&48.6&39.6&36.6&2.4&1.2&30.0&1.8&6.0\\
1-4&0.0&19.2&0.0&15.6&0.0&0.0&0.0&0.0&0.0&0.0&0.0&0.0&1.2&0.0&0.0\\
2-3&100&100&100&100&100&100&100&100&100&100&100&100&100&100&100\\
2-4&1.2&100&0.0&100&0.0&1.2&0.0&2.4&2.4&0.6&0.6&0.6&17.4&0.6&0.0\\
3-4&6.0&100&3.6&100&3.0&8.4&8.4&12.6&10.2&7.8&28.2&19.2&64.2&24.6&33.6\\
\noalign{\smallskip}\hline\noalign{\smallskip}
\end{tabular}
%\end{scriptsize}
\end{table}
}

\subsection{Simulation study based on DTI data set}
To check the correctness of the above results for the DTI data set, we conduct an additional simulation study based on these data. To mimic the observation given in the data example, we generated the simulation data in the following way: 
\begin{itemize}
\item we used $\ell=4$ repeated samples with $n=17$ functional observations at $93$ design time points;
\item the covariance function $\gamma(s,t)$, $s,t\in[0,\ell]$ was equal to the sample covariance function~\eqref{cov_fun_est} for the given data set;
\item for checking the type I error control, in each group, the mean function was equal to the
sample mean function for the pooled sample of $\ell\cdot n=4\cdot 17=68$ observations collected from four original samples;
\item for the investigation of power, the mean function in the $i$-th repeated sample was equal to the sample mean function for the $i$-th sample from the data set, $i=1,\dots,\ell$.
\end{itemize}
For given values of $\ell$ and $n$, the $93$-dimensional data were generated from the normal ($N$), Student ($t_5$), and chi-square ($\chi_{10}^2$) distributions having expected value and covariance matrix equal to the above sample mean functions and sample covariance function respectively. Thus, we consider symmetric, heavily tailed, and skewed distributions. 

In the case of the global null hypothesis (Table~\ref{table_srde_1}), we observe that the $\mathcal{P}_2$ and $\mathcal{B}_2$ $\mathcal{C}_{17}(4)$ tests are extremely conservative, which was also noted in the simulations of Section~\ref{sec_4} for larger correlation. This conservativeness results in significant loss of power for these tests -- the largest empirical power among them is $20.8\%$, while the remaining testing procedures have power equal to $100\%$ or slightly smaller. These facts explain the testing results for the global null hypothesis presented in Section~\ref{sec_5_1}. Note, however, that some of the tests do not control the type I error level, namely $\mathcal{P}_1$ $\mathcal{C}_{17}(4)$; $\mathcal{P}_1$ $\mathcal{D}_{17}(4)$; $\mathcal{P}_1$, $\mathcal{P}_2$, and $\mathcal{B}_2$ $\mathcal{E}_{17}(4)$. This is due to the use of a smaller number of observations ($n=17$) and four groups ($\ell=4$). 

Let us now turn to the pairwise comparisons (Table~\ref{table_srde_2}). All tests control the FWER at level $5\%$, but the following tests have a conservative character: $\mathcal{P}_2$ and $\mathcal{B}_2$ $\mathcal{C}_{17}(4)$; $\mathcal{P}_2$, $\mathcal{B}_1$, and $\mathcal{B}_2$ $\mathcal{D}_{17}(4)$; $\mathcal{B}_1$ $\mathcal{E}_{17}(4)$. Once again, the first two of these are extremely conservative. This usually results in a lower power compared with the other tests. These observations are consequences of applying the tests to just two groups and using the Bonferroni correction. For the power investigation, we consider six cases corresponding to the six pairwise comparisons. The empirical power varies for different comparisons. The largest powers appear for 1-4 and 2-4, which explains the detection of significant differences for these comparisons in the analysis of the DTI data set (Section~\ref{sec_5_1}). On the other hand, for 2-3, all tests have a smaller power, which is a result of the fact that visits 2 and 3 have the most similar FA profiles. For 1-3 and 3-4, the $\mathcal{P}_1$, $\mathcal{P}_2$, and $\mathcal{B}_2$ $\mathcal{E}_{17}(4)$ tests and the $\mathcal{B}_1$ and $\mathcal{B}_3$ $\mathcal{C}_{17}(4)$ tests respectively have the largest power; this explains the rejections of equality of mean functions in these cases noted in Section~\ref{sec_5_1}.

{\small\tabcolsep=5.5pt
\begin{table}[!t]%[!ht]
\caption[]{Empirical sizes and powers (as percentages) of all tests for the global null hypothesis~\eqref{H0fda} obtained in the simulation study based on the DTI data set.}
\label{table_srde_1}
%\begin{scriptsize}
\centering
\begin{tabular}{l rrrrr rrrrr rrrrr}
\hline\noalign{\smallskip}
&\multicolumn{5}{l}{$\mathcal{C}_{17}(4)$}  & \multicolumn{5}{l}{$\mathcal{D}_{17}(4)$}  & \multicolumn{5}{l}{$\mathcal{E}_{17}(4)$}\\%\cline{2-6}\cline{7-11}\cline{12-16}
Distr& $\mathcal{P}_1$&$\mathcal{P}_2$&$\mathcal{B}_1$&$\mathcal{B}_2$&$\mathcal{B}_3$  & $\mathcal{P}_1$&$\mathcal{P}_2$&$\mathcal{B}_1$&$\mathcal{B}_2$&$\mathcal{B}_3$  &  $\mathcal{P}_1$&$\mathcal{P}_2$&$\mathcal{B}_1$&$\mathcal{B}_2$&$\mathcal{B}_3$\\
\hline\noalign{\smallskip}
%\noalign{\smallskip}\noalign{\smallskip}
\multicolumn{5}{l}{Empirical sizes} &  &&&&  &  &&&&\\
$N$&9.1&0.0&4.2&0.0&3.8&8.9&5.6&2.4&4.5&3.3&9.1&12.0&2.1&9.0&4.1\\
$t_5$&10.9&0.0&3.4&0.0&3.5&11.3&6.1&1.9&5.0&3.2&11.2&13.5&2.1&9.1&3.7\\
$\chi_{10}^2$&9.6&0.0&4.9&0.0&4.3&9.8&6.1&2.9&5.4&3.7&8.6&10.7&3.1&8.3&4.2\\
\multicolumn{5}{l}{Empirical powers} &  &&&&  &  &&&&\\
$N$&100&8.1&100&11.4&100&100&100&99.9&100&100&100&100&100&100&100\\
$t_5$&99.8&17.8&99.2&20.8&99.0&100&99.8&97.5&99.7&99.1&100&100&98.6&100&99.9\\
$\chi_{10}^2$&100&8.4&100&11.1&99.9&100&100&99.8&99.9&99.9&100&100&99.9&100&100\\
\noalign{\smallskip}\hline\noalign{\smallskip}
\end{tabular}
%\end{scriptsize}
\end{table}
}

{\small\tabcolsep=5.5pt
\begin{table}[!t]%[!ht]
\caption[]{Empirical FWER and powers (as percentages) of all tests for the pairwise comparisons obtained in the simulation study based on the DTI data set.}
\label{table_srde_2}
%\begin{scriptsize}
\centering
\begin{tabular}{l rrrrr rrrrr rrrrr}
\hline\noalign{\smallskip}
&\multicolumn{5}{l}{$\mathcal{C}_{17}(4)$}  & \multicolumn{5}{l}{$\mathcal{D}_{17}(4)$}  & \multicolumn{5}{l}{$\mathcal{E}_{17}(4)$}\\%\cline{2-6}\cline{7-11}\cline{12-16}
& $\mathcal{P}_1$&$\mathcal{P}_2$&$\mathcal{B}_1$&$\mathcal{B}_2$&$\mathcal{B}_3$  & $\mathcal{P}_1$&$\mathcal{P}_2$&$\mathcal{B}_1$&$\mathcal{B}_2$&$\mathcal{B}_3$  &  $\mathcal{P}_1$&$\mathcal{P}_2$&$\mathcal{B}_1$&$\mathcal{B}_2$&$\mathcal{B}_3$\\
\hline\noalign{\smallskip}
\multicolumn{6}{l}{Empirical FWER} &  &&&&  &  &&&&\\
$N$&4.7&0.0&6.1&0.0&5.2&3.7&1.6&2.3&1.3&4.4&3.8&5.8&0.6&4.8&4.3\\
$t_5$&4.7&0.0&4.5&0.0&4.5&4.2&1.2&1.0&1.4&3.5&4.0&5.2&0.3&3.2&2.6\\
$\chi_{10}^2$&4.2&0.0&5.3&0.0&5.1&2.8&0.7&1.5&0.8&3.9&4.2&6.6&0.5&4.7&4.5\\
\multicolumn{6}{l}{Empirical powers} &  &&&&  &  &&&&\\
\multicolumn{6}{l}{$N$} &  &&&&  &  &&&&\\
1-2&55.4&0.0&55.3&0.0&53.7&59.0&37.0&36.8&33.1&53.3&56.2&63.5&18.7&59.8&57.6\\
1-3&30.4&0.0&31.4&0.1&30.5&33.6&23.3&19.3&19.9&30.6&69.3&74.7&35.5&69.7&72.2\\
1-4&100&20.7&100&28.5&100&100&99.9&99.9&99.9&100&100&100&97.8&100&100\\
2-3&3.5&0.0&4.0&0.0&3.8&3.5&1.1&1.2&1.1&3.4&9.9&12.6&1.4&11.6&10.3\\
2-4&90.6&0.8&92.4&1.1&91.4&92.6&85.0&82.5&82.6&90.8&96.9&98.0&73.2&97.4&97.7\\
3-4&59.3&0.0&67.4&0.0&66.9&57.0&48.4&48.7&46.9&60.5&56.6&62.8&21.2&56.9&56.6\\
\multicolumn{6}{l}{$t_5$} &  &&&&  &  &&&&\\
1-2&64.6&0.0&54.8&0.0&56.3&71.2&47.4&42.0&42.7&60.7&75.4&78.4&23.9&71.7&71.2\\
1-3&38.4&0.2&34.2&0.3&34.0&48.6&33.7&25.0&27.1&39.3&80.7&82.7&44.1&77.0&79.2\\
1-4&99.0&33.4&98.4&35.8&98.7&100&99.7&98.3&99.2&99.8&99.9&99.9&94.4&99.9&99.9\\
2-3&5.2&0.0&3.9&0.0&3.4&5.7&2.2&1.4&2.0&3.6&19.9&23.4&3.6&18.1&16.4\\
2-4&92.7&1.5&88.7&2.8&90.8&95.0&89.8&83.9&87.3&91.8&98.4&99.2&74.1&97.1&97.6\\
3-4&64.5&0.2&68.5&0.7&68.4&65.4&59.4&52.9&55.6&64.6&70.6&74.1&26.1&67.0&66.3\\
\multicolumn{6}{l}{$\chi_{10}^2$} &  &&&&  &  &&&&\\
1-2&58.5&0.0&58.2&0.0&57.1&62.1&37.6&40.3&34.2&55.1&58.6&67.6&19.3&63.0&60.9\\
1-3&32.2&0.1&34.2&0.2&32.4&39.1&26.5&20.9&22.1&35.3&73.6&77.7&38.4&73.7&74.7\\
1-4&100&20.2&100&28.8&100&100&99.8&99.9&99.9&100&100&100&97.7&100&99.9\\
2-3&4.3&0.0&4.6&0.0&4.6&4.4&2.1&1.8&1.7&4.0&12.5&16.0&2.6&14.0&12.8\\
2-4&93.4&0.8&93.8&1.0&93.9&94.7&89.2&85.4&86.4&93.6&98.4&98.9&77.5&98.0&98.4\\
3-4&58.5&0.0&68.7&0.1&67.5&56.4&50.4&49.5&47.8&60.5&58.2&64.4&22.6&60.1&58.9\\
\noalign{\smallskip}\hline\noalign{\smallskip}
\end{tabular}
%\end{scriptsize}
\end{table}
}

\section{Conclusions and future work}
\label{sec_6}
We have considered one-way repeated measures ANOVA for functional data, which aims to find significant differences in mean functions representing successive experimental conditions applied to the same subjects. Despite the evident interest in this analysis, it is rarely considered in the literature. We have investigated testing procedures for any number of repeated samples, not just two as in most previous studies. Moreover, the procedures take into account both mean and variance, which usually results in more powerful tests. As well as procedures for testing the global null hypothesis, we also propose a simple post hoc method, which is of great interest since practitioners usually wish to know in which cases the significant differences appear, not just that such differences exist. To approximate the null distribution of test statistics, two permutation and three bootstrap methods were used. These are distribution-free methods, which have achieved promising levels of performance in previous studies. However, for the present problem, they do not perform equally well, and it is not easy to suggest the best test for each case. The simulation results nevertheless indicated that the parametric bootstrap $\mathcal{E}_n(\ell)$ test seems to have very good properties in terms of type I error level control and power. This test also usually performs much better than the procedures based on the test statistic of Mart\'inez-Camblor and Corral (2011). Nevertheless, the choice of the best test for a given practical problem can be assisted by a simulation study based on available data, as we showed in Section~\ref{sec_5}, where we applied the tests under consideration to an important medical problem.

There are many possible directions for future work. One is to extend the methodology to more complicated designs than the one-way classification. A second direction is the development of more sophisticated post hoc tests, which may be more powerful than our simple procedure. These and other problems will be addressed in future research.

\section*{Acknowledgements}

A part of the calculations for the simulation study was carried out at the Pozna\'n Supercomputing and Networking Center (grant no. 617).

\noindent \begin{center}\begin{large}{\bf REFERENCES}\end{large}\end{center}

\everypar = {
\parindent=0pt
\hangindent=8mm
\hangafter=1
}\noindent 

AMRO, L., PAULY, M., RAMOSAJ, B., (2021). Asymptotic‐Based Bootstrap Approach for Matched Pairs with Missingness in a Single Arm. \textit{Biometrical Journal}, 63, 1389--1405.

BOX, G. E. P., (1954). Some Theorems on Quadratic Forms Applied in the Study of Analysis of Variance Problems, I. Effect of Inequality of Variance in the One-Way Classification. \textit{The Annals of Mathematical Statistics}, 25, 290--302.

%Cuevas, A., Febrero, M., Fraiman, R.: An ANOVA test for functional data. Comput. Statist. Data Anal. \textbf{47}, 111--122 (2004)
CUEVAS, A., FEBRERO, M., FRAIMAN, R., (2004). An ANOVA Test for Functional Data. \textit{Computational Statistics \& Data Analysis}, 47, 111--122.

%Ditzhaus, M., Genuneit, J., Janssen, A., Pauly, M. (2021). CASANOVA: Permutation inference in factorial survival designs. Biometrics, 1--13.
DITZHAUS, M., GENUNEIT, J., JANSSEN, A., PAULY, M., (2021). CASANOVA: Permutation Inference in Factorial Survival Designs. \textit{Biometrics}, 1--13.

%Ferraty, F., Vieu, P.: Nonparametric Functional Data Analysis: Theory and Practice. Springer, New York (2006)
FERRATY, F., VIEU, P., (2006). \textit{Nonparametric Functional Data Analysis: Theory and Practice}. New York: Springer.  

%Goldsmith, J., Scheipl, F., Huang, L., Wrobel, J., Di, C., Gellar, J., Harezlak, J., McLean, M.W., Swihart, B., Xiao, L., Crainiceanu, C., Reiss, P.T. (2022). refund: Regression with Functional Data. R package version 0.1-28, \url{https://CRAN.R-project.org/package=refund}.
GOLDSMITH, J., SCHEIPL, F., HUANG, L., WROBEL, J., DI, C., GELLAR, J., HAREZLAK, J., MCLEAN, M.W., SWIHART, B., XIAO, L., CRAINICEANU, C., REISS, P. T., (2022). refund: Regression with Functional Data. R package version 0.1-28, \url{https://CRAN.R-project.org/package=refund}.

%Horv\'ath, L., Kokoszka, P.: Inference for Functional Data with Applications. Springer, New York (2012)
HORV\'ATH, L., KOKOSZKA, P., (2012). \textit{Inference for Functional Data with Applications}. New York: Springer.

%Konietschke, F., Bathke, A.C., Harrar, S.W., Pauly, M.: Parametric and nonparametric bootstrap methods for general MANOVA. J. Multivariate Anal. \textbf{140}, 291--301 (2015)
KONIETSCHKE, F., BATHKE, A. C., HARRAR, S. W., PAULY, M., (2015). Parametric and Nonparametric Bootstrap Methods for General MANOVA.  \textit{Journal of Multivariate Analysis}, 140, 291--301.

%Konietschke, F., Pauly, M.: Bootstrapping and permuting paired $t$-test type statistics. Stat. Comput. \textbf{24}, 283--296 (2014)
KONIETSCHKE, F., PAULY, M., (2014). Bootstrapping and Permuting Paired $t$-Test Type Statistics. \textit{Statistics and Computing}, 24, 283--296.

%MART\'INEZ-CAMBLOR, P., CORRAL, N., (2011). Repeated measures analysis for functional data. \textit{Computational Statistics and Data Analysis}, 55, 3244--3256.
MART\'INEZ-CAMBLOR, P., CORRAL, N., (2011). Repeated Measures Analysis for Functional Data. \textit{Computational Statistics \& Data Analysis}, 55, 3244--3256.

MRKVI\v{C}KA, T., MYLLYM\"{A}KI, M., J\'ILEK, M., HAHN, U., (2020). A One-Way ANOVA Test for Functional Data with Graphical Interpretation. \textit{Kybernetika}, 56, 432--458.

%Pini, A., Vantini, S., Colosimo, B. M., and Grasso, M. (2018). Domain-selective functional analysis of variance for supervised statistical profile monitoring of signal data. Journal of the Royal Statistical Society: Series C (Applied Statistics), 67, 55--81.
PINI, A., VANTINI, S., COLOSIMO, B. M., GRASSO, M., (2018). Domain-Selective Functional Analysis of Variance for Supervised Statistical Profile Monitoring of Signal Data. \textit{Journal of the Royal Statistical Society: Series C (Applied Statistics)}, 67, 55--81.

%Ramsay, J.O., Hooker, G., Graves, S., (2009). Functional Data Analysis with R and MATLAB. Springer, Berlin.
RAMSAY, J. O., HOOKER, G., GRAVES, S., (2009). \textit{Functional Data Analysis with R and MATLAB}. Berlin: Springer.

%Ramsay, J.O., Silverman, B.W.: Functional Data Analysis, 2nd Edition. Springer, New York (2005)
RAMSAY, J. O., SILVERMAN, B.W., (2005). \textit{Functional Data Analysis}, 2nd Edition. New York: Springer.

R CORE TEAM, (2022). R: A Language and Environment for Statistical Computing. R Foundation for Statistical Computing, Vienna, Austria. \url{https://www.R-project.org/}

SMAGA, \L., (2019). Repeated Measures Analysis for Functional Data Using Box-Type Approximation -- with Applications. \textit{REVSTAT -- Statistical Journal}, 17, 523--549.

SMAGA, \L., (2020). A Note on Repeated Measures Analysis for Functional Data. \textit{AStA Advances in Statistical Analysis}, 104, 117--139.

%SMAGA, \L., (2021). One-Way Repeated Measures ANOVA for Functional Data. In: \textit{Data Analysis and Rationality in a Complex World. IFCS 2019. Studies in Classification, Data Analysis, and Knowledge Organization}. T. Chadjipadelis, B. Lausen, A. Markos, T. R. Lee, A. Montanari, R. Nugent (eds): Springer, Cham. 243--251.
SMAGA, \L., (2021). One-Way Repeated Measures ANOVA for Functional Data. In: \textit{Data Analysis and Rationality in a Complex World. IFCS 2019. Studies in Classification, Data Analysis, and Knowledge Organization}. T. Chadjipadelis, B. Lausen, A. Markos, T. R. Lee, A. Montanari, R. Nugent (eds.): Springer, Cham. 243--251.

SMAGA, \L., ZHANG, J. T., (2020). Linear Hypothesis Testing for Weighted Functional Data with Applications. \textit{Scandinavian Journal of Statistics}, 47, 493--515.

TUKEY, J. W., (1953). The Problem of Multiple Comparisons. Princeton University.

%Zhang, J. T.: Analysis of Variance for Functional Data. Chapman \& Hall, London (2013)
ZHANG, J. T., (2013). \textit{Analysis of Variance for Functional Data}. London: Chapman \& Hall.

%Zhang, J. T., Cheng, M. Y., Wu, H. T., Zhou, B.: A new test for functional one-way ANOVA with applications to ischemic heart screening. Comput. Statist. Data Anal. \textbf{132}, 3--17 (2019)

ZHANG, J. T., CHENG, M. Y., WU, H. T., ZHOU, B., (2019). A New Test for Functional One-Way ANOVA with Applications to Ischemic Heart Screening. \textit{Computational Statistics \& Data Analysis}, 132, 3--17.

\newpage

	\begin{center} 
		{\Large \bf Supplementary materials to \\ Functional repeated measures analysis of variance and its application} 
	\end{center}

\vspace*{-4mm}
\begin{center}
	\begin{large} 
	{\bf Katarzyna Kury\l o}\footnote{Faculty of Mathematics and Computer Science, Adam Mickiewicz University, Pozna\'{n}, Poland. E-mail: katkur7@st.amu.edu.pl},
	{\bf \L ukasz Smaga}\footnote{Faculty of Mathematics and Computer Science, Adam Mickiewicz University, Pozna\'{n}, Poland. E-mail: ls@amu.edu.pl, Corresponding author}	 
    \end{large}
\end{center}

In this supporting material, we present all results of the simulation study of Section~4 of the main paper. They are depicted in Tables~\ref{table1_sm}-\ref{table16_sm}.
\listoftables
\smallskip

\clearpage

\setcounter{table}{0}

\begin{table}[!t]
\caption[Empirical sizes of all tests obtained in model~M1]{Empirical sizes (as percentages) of all tests obtained in model~M1.}
\label{table1_sm}
\centering
%\begin{scriptsize}
\begin{tabular}{l rrrrr rrrrr rrrrr}
\hline\noalign{\smallskip}
& \multicolumn{5}{l}{$\mathcal{C}_n(3)$} &  \multicolumn{5}{l}{$\mathcal{D}_n(3)$} &   \multicolumn{5}{l}{$\mathcal{E}_n(3)$}\\%\cline{3-7}\cline{9-13}\cline{15-19}
$\rho$& $\mathcal{P}_1$&$\mathcal{P}_2$&$\mathcal{B}_1$&$\mathcal{B}_2$&$\mathcal{B}_3$  &  $\mathcal{P}_1$&$\mathcal{P}_2$&$\mathcal{B}_1$&$\mathcal{B}_2$&$\mathcal{B}_3$  &  $\mathcal{P}_1$&$\mathcal{P}_2$&$\mathcal{B}_1$&$\mathcal{B}_2$&$\mathcal{B}_3$\\
\hline\noalign{\smallskip}
%\noalign{\smallskip}\noalign{\smallskip}
&\multicolumn{2}{l}{normal}&&  &  &&&&  &  &&&&\\
&\multicolumn{2}{l}{$n=35$}&&  &  &&&&  &  &&&&\\
0.00&  4.9&5.4&5.0&5.5&4.9 & 5.1&5.5&3.6&4.4&4.5 & 4.1&4.2&2.9&3.4&3.3\\
0.25&  4.8&1.7&4.8&2.1&4.6 & 5.7&5.3&3.4&4.6&4.0 & 6.1&6.1&3.4&4.4&3.7\\
0.50&  5.3&0.5&4.8&0.8&4.5 & 5.2&4.5&3.7&4.7&4.0 & 6.7&7.1&3.7&6.0&3.9\\
0.75&  4.7&0.0&3.9&0.0&4.2 & 5.6&5.2&3.3&5.1&4.0 & 8.2&8.1&3.7&7.4&4.3\\
\end{tabular}
%\end{scriptsize}
\end{table}

{\small\tabcolsep=4.7pt
\begin{table}[!t]
\caption[Empirical powers of all tests obtained in model~M2]{Empirical powers (as percentages) of all tests obtained in model~M2.}
\label{table2_sm}
\centering
%\begin{scriptsize}
\begin{tabular}{l rrrrr rrrrr rrrrr}%{p{0.3cm}p{0.11cm}p{0.62cm}p{0.62cm}p{0.62cm}p{0.62cm}p{0.62cm}p{0.02cm}p{0.62cm}p{0.62cm}p{0.62cm}p{0.62cm}p{0.62cm}p{0.02cm}p{0.62cm}p{0.62cm}p{0.62cm}p{0.62cm}p{0.62cm}}
\hline\noalign{\smallskip}
& \multicolumn{5}{l}{$\mathcal{C}_n(3)$}  & \multicolumn{5}{l}{$\mathcal{D}_n(3)$}  & \multicolumn{5}{l}{$\mathcal{E}_n(3)$}\\%\cline{2-6}\cline{7-11}\cline{12-16}
$\rho$&  $\mathcal{P}_1$&$\mathcal{P}_2$&$\mathcal{B}_1$&$\mathcal{B}_2$&$\mathcal{B}_3$  & $\mathcal{P}_1$&$\mathcal{P}_2$&$\mathcal{B}_1$&$\mathcal{B}_2$&$\mathcal{B}_3$  &  $\mathcal{P}_1$&$\mathcal{P}_2$&$\mathcal{B}_1$&$\mathcal{B}_2$&$\mathcal{B}_3$\\
\hline\noalign{\smallskip}
%\noalign{\smallskip}\noalign{\smallskip}
&\multicolumn{2}{l}{normal}&&  &  &&&&  &  &&&&\\
&\multicolumn{2}{l}{$n=35$}&&  &  &&&&  &  &&&&\\
0.00& 32.5&32.1&30.6&32.0&30.0 & 45.0&45.1&34.8&38.4&38.5 & 88.7&88.3&84.4&85.8&84.5\\
0.25& 42.9&24.7&40.6&24.5&40.2 & 59.5&59.0&47.6&54.0&52.1 & 96.1&95.9&92.1&94.8&93.3\\
0.50& 65.6&13.7&59.3&14.8&59.6 & 82.1&81.9&72.3&78.8&74.0 & 99.9&100&98.9&99.7&99.4\\
0.75& 97.8&5.3&96.6&5.7&96.2 & 99.9&99.8&98.7&99.7&99.5 & 100&100&100&100&100\\
\end{tabular}
%\end{scriptsize}
\end{table}
}

{\small\tabcolsep=4.7pt
\clearpage
\begin{table}[!t]
\caption[Empirical powers of all tests obtained in model~M3]{Empirical powers (as percentages) of all tests obtained in model~M3.}
\label{table3_sm}
\centering
%\begin{scriptsize}
\begin{tabular}{l rrrrr rrrrr rrrrr}
\hline\noalign{\smallskip}
& \multicolumn{5}{l}{$\mathcal{C}_n(3)$}  & \multicolumn{5}{l}{$\mathcal{D}_n(3)$}  & \multicolumn{5}{l}{$\mathcal{E}_n(3)$}\\%\cline{2-6}\cline{7-11}\cline{12-16}
$\rho$&  $\mathcal{P}_1$&$\mathcal{P}_2$&$\mathcal{B}_1$&$\mathcal{B}_2$&$\mathcal{B}_3$  & $\mathcal{P}_1$&$\mathcal{P}_2$&$\mathcal{B}_1$&$\mathcal{B}_2$&$\mathcal{B}_3$  &  $\mathcal{P}_1$&$\mathcal{P}_2$&$\mathcal{B}_1$&$\mathcal{B}_2$&$\mathcal{B}_3$\\
\hline\noalign{\smallskip}
&\multicolumn{2}{l}{normal}&&  &  &&&&  &  &&&&\\
&\multicolumn{2}{l}{$n=35$}&&  &  &&&&  &  &&&&\\
0.00 & 92.9 & 93.6 & 91.1 & 93.1 & 91.5 & 99.3 & 99.7 & 98.2 & 98.9 & 99.2
& 100 & 100 & 100 & 100 & 100 \\
0.25 & 98.3 & 91.3 & 97.6 & 91.2 & 97.5 & 99.8 & 99.8 & 99.4 & 99.8 &
99.7 & 100 & 100 & 100 & 100 & 100 \\
0.50 & 99.8 & 90.9 & 99.7 & 91.0 & 99.8 & 100 & 100 & 100 & 100 & 100 & 100
& 100 & 100 & 100 & 100 \\
0.75 & 100 & 88.4 & 100 & 89.3 & 100 & 100 & 100 & 100 & 100 & 100 & 100
& 100 & 100 & 100 & 100 \\
\end{tabular}
%\end{scriptsize}
\end{table}
}

\clearpage
\begin{table}[!t]
\caption[Empirical sizes of all tests obtained in model~M4]{Empirical sizes (as percentages) of all tests obtained in model~M4.}
\label{table4_sm}
\centering
%\begin{scriptsize}
\begin{tabular}{l rrrrr rrrrr rrrrr}
\hline\noalign{\smallskip}
& \multicolumn{5}{l}{$\mathcal{C}_n(3)$}  & \multicolumn{5}{l}{$\mathcal{D}_n(3)$}  & \multicolumn{5}{l}{$\mathcal{E}_n(3)$}\\%\cline{2-6}\cline{7-11}\cline{12-16}
$\rho$&  $\mathcal{P}_1$&$\mathcal{P}_2$&$\mathcal{B}_1$&$\mathcal{B}_2$&$\mathcal{B}_3$  & $\mathcal{P}_1$&$\mathcal{P}_2$&$\mathcal{B}_1$&$\mathcal{B}_2$&$\mathcal{B}_3$  &  $\mathcal{P}_1$&$\mathcal{P}_2$&$\mathcal{B}_1$&$\mathcal{B}_2$&$\mathcal{B}_3$\\
\hline\noalign{\smallskip}
&\multicolumn{2}{l}{normal}&&  &  &&&&  &  &&&&\\
&\multicolumn{2}{l}{$n=35$}&&  &  &&&&  &  &&&&\\
0.00 & 5.9 & 6.1 & 5.7 & 6.2 & 5.8 & 5.4 & 5.2 & 4.0 & 4.5 & 4.6 & 5.7 & 5.6
& 3.6 & 4.3 & 3.7 \\
0.25 & 6.0 & 2.4 & 6.0 & 2.8 & 5.7 & 5.2 & 4.9 & 3.8 & 4.5 & 3.8 & 5.6 & 5.7
& 2.9 & 4.9 & 3.1 \\
0.50 & 4.5 & 0.7 & 4.6 & 0.9 & 4.3 & 5.0 & 5.0 & 3.9 & 4.9 & 4.1 & 6.9 & 6.8
& 3.2 & 5.5 & 3.5 \\
0.75 & 6.3 & 0.1 & 6.2 & 0.1 & 6.0 & 6.4 & 6.3 & 5.3 & 6.3 & 5.4 & 8.4 & 8.7
& 3.0 & 7.6 & 4.0 \\
\end{tabular}
%\end{scriptsize}
\end{table}

\clearpage
{\small\tabcolsep=4.7pt
\begin{table}[!t]
\caption[Empirical powers of all tests obtained in model~M5]{Empirical powers (as percentages) of all tests obtained in model~M5.}
\label{table5_sm}
\centering
%\begin{scriptsize}
\begin{tabular}{l rrrrr rrrrr rrrrr}
\hline\noalign{\smallskip}
& \multicolumn{5}{l}{$\mathcal{C}_n(3)$}  & \multicolumn{5}{l}{$\mathcal{D}_n(3)$}  & \multicolumn{5}{l}{$\mathcal{E}_n(3)$}\\%\cline{2-6}\cline{7-11}\cline{12-16}
$\rho$&  $\mathcal{P}_1$&$\mathcal{P}_2$&$\mathcal{B}_1$&$\mathcal{B}_2$&$\mathcal{B}_3$  & $\mathcal{P}_1$&$\mathcal{P}_2$&$\mathcal{B}_1$&$\mathcal{B}_2$&$\mathcal{B}_3$  &  $\mathcal{P}_1$&$\mathcal{P}_2$&$\mathcal{B}_1$&$\mathcal{B}_2$&$\mathcal{B}_3$\\
\hline\noalign{\smallskip}
&\multicolumn{2}{l}{normal}&&  &  &&&&  &  &&&&\\
&\multicolumn{2}{l}{$n=35$}&&  &  &&&&  &  &&&&\\
0.00 & 55.1 & 54.4 & 54.1 & 54.2 & 54.1 & 63.2 & 63.4 & 55.2 & 60.0 & 60.0 & 91.0
& 91.1 & 86.7 & 88.7 & 88.2 \\
0.25 & 61.3 & 47.5 & 61.8 & 49.2 & 61.8 & 72.4 & 72.2 & 66.2 & 70.6 &
69.1 & 96.6 & 96.8 & 94.1 & 96.2 & 93.8 \\
0.50 & 77.6 & 40.6 & 76.7 & 42.2 & 75.8 & 89.0 & 89.2 & 83.0 & 87.5 & 85.6 &
98.9 & 99.0 & 97.9 & 98.6 & 98.3 \\
0.75 & 97.3 & 34.3 & 97.0 & 37.8 & 97.1 & 99.8 & 99.8 & 99.4 & 99.8 & 99.6
& 100 & 100 & 100 & 100 & 100 \\
\end{tabular}
%\end{scriptsize}
\end{table}
}

\clearpage
{\small\tabcolsep=4.7pt
\begin{table}[!t]
\caption[Empirical powers of all tests obtained in model~M6]{Empirical powers (as percentages) of all tests obtained in model~M6.}
\label{table6_sm}
\centering
%\begin{scriptsize}
\begin{tabular}{l rrrrr rrrrr rrrrr}
\hline\noalign{\smallskip}
& \multicolumn{5}{l}{$\mathcal{C}_n(3)$}  & \multicolumn{5}{l}{$\mathcal{D}_n(3)$}  & \multicolumn{5}{l}{$\mathcal{E}_n(3)$}\\%\cline{2-6}\cline{7-11}\cline{12-16}
$\rho$&  $\mathcal{P}_1$&$\mathcal{P}_2$&$\mathcal{B}_1$&$\mathcal{B}_2$&$\mathcal{B}_3$  & $\mathcal{P}_1$&$\mathcal{P}_2$&$\mathcal{B}_1$&$\mathcal{B}_2$&$\mathcal{B}_3$  &  $\mathcal{P}_1$&$\mathcal{P}_2$&$\mathcal{B}_1$&$\mathcal{B}_2$&$\mathcal{B}_3$\\
\hline\noalign{\smallskip}
&\multicolumn{2}{l}{normal}&&  &  &&&&  &  &&&&\\
&\multicolumn{2}{l}{$n=35$}&&  &  &&&&  &  &&&&\\
0.00 & 67.5 & 68.4 & 67.1 & 68.3 & 67.5 & 75.4 & 75.9 & 70.7 & 72.9 & 72.6
& 96.7 & 96.4 & 93.7 & 95.1 & 94.5 \\
0.25 & 71.9 & 60.7 & 72.5 & 62.9 & 71.9 & 82.1 & 82.3 & 77.2 & 80.7 &
79.4 & 98.6 & 99.0 & 98.0 & 98.4 & 97.8 \\
0.50 & 87.5 & 57.9 & 86.2 & 60.7 & 86.5 & 95.2 & 94.8 & 92.0 & 94.0 & 93.2 &
100 & 99.9 & 99.6 & 99.9 & 99.9 \\
0.75 & 98.8 & 52.7 & 98.6 & 56.2 & 98.9 & 100 & 100 & 100 & 100 & 100 &
100 & 100 & 100 & 100 & 100 \\
\end{tabular}
%\end{scriptsize}
\end{table}
}

%{\small\tabcolsep=4.7pt
\begin{table}[!t]
\caption[Empirical FWER of all post hoc tests obtained in model~M1]{Empirical FWER (as percentages) of all post hoc tests obtained in model~M1.}
\label{table7_sm}
\centering
%\begin{scriptsize}
\begin{tabular}{l rrrrr rrrrr rrrrr}%{p{0.3cm}p{0.11cm}p{0.62cm}p{0.62cm}p{0.62cm}p{0.62cm}p{0.62cm}p{0.02cm}p{0.62cm}p{0.62cm}p{0.62cm}p{0.62cm}p{0.62cm}p{0.02cm}p{0.62cm}p{0.62cm}p{0.62cm}p{0.62cm}p{0.62cm}}
\hline\noalign{\smallskip}
& \multicolumn{5}{l}{$\mathcal{C}_n(3)$}  & \multicolumn{5}{l}{$\mathcal{D}_n(3)$}  & \multicolumn{5}{l}{$\mathcal{E}_n(3)$}\\%\cline{2-6}\cline{7-11}\cline{12-16}
$\rho$&  $\mathcal{P}_1$&$\mathcal{P}_2$&$\mathcal{B}_1$&$\mathcal{B}_2$&$\mathcal{B}_3$  & $\mathcal{P}_1$&$\mathcal{P}_2$&$\mathcal{B}_1$&$\mathcal{B}_2$&$\mathcal{B}_3$  &  $\mathcal{P}_1$&$\mathcal{P}_2$&$\mathcal{B}_1$&$\mathcal{B}_2$&$\mathcal{B}_3$\\
\hline\noalign{\smallskip}
&\multicolumn{2}{l}{normal}&&  &  &&&&  &  &&&&\\
&\multicolumn{2}{l}{$n=35$}&&  &  &&&&  &  &&&&\\
0.00 & 4.9 & 4.7 & 5.9 & 5.3 & 5.5 & 4.5 & 4.5 & 4.0 & 3.9 & 4.6 & 5.3 & 5.4
& 3.5 & 4.6 & 5.6 \\
0.25 & 4.4 & 2.3 & 5.4 & 2.4 & 4.9 & 4.2 & 3.8 & 3.3 & 3.6 & 3.6 & 4.1 &
4.3 & 2.4 & 3.6 & 4.1 \\
0.50 & 4.0 & 0.5 & 4.9 & 0.8 & 4.9 & 3.8 & 3.7 & 3.3 & 3.7 & 3.9 & 3.8 & 3.8
& 2.5 & 3.7 & 3.7 \\
0.75 & 5.4 & 0.1 & 6.4 & 0.1 & 6.4 & 5.2 & 5.9 & 4.5 & 4.8 & 5.6 & 5.4 &
5.3 & 3.3 & 5.1 & 5.0 \\
\end{tabular}
%\end{scriptsize}
\end{table}
%}

\clearpage
\begin{table}[!t]
\caption[Empirical FWER of all post hoc tests obtained in model~M4]{Empirical FWER (as percentages) of all post hoc tests obtained in model~M4.}
\label{table8_sm}
\centering
%\begin{scriptsize}
\begin{tabular}{l rrrrr rrrrr rrrrr}
\hline\noalign{\smallskip}
& \multicolumn{5}{l}{$\mathcal{C}_n(3)$}  & \multicolumn{5}{l}{$\mathcal{D}_n(3)$}  & \multicolumn{5}{l}{$\mathcal{E}_n(3)$}\\%\cline{2-6}\cline{7-11}\cline{12-16}
$\rho$&  $\mathcal{P}_1$&$\mathcal{P}_2$&$\mathcal{B}_1$&$\mathcal{B}_2$&$\mathcal{B}_3$  & $\mathcal{P}_1$&$\mathcal{P}_2$&$\mathcal{B}_1$&$\mathcal{B}_2$&$\mathcal{B}_3$  &  $\mathcal{P}_1$&$\mathcal{P}_2$&$\mathcal{B}_1$&$\mathcal{B}_2$&$\mathcal{B}_3$\\
\hline\noalign{\smallskip}
&\multicolumn{2}{l}{normal}&&  &  &&&&  &  &&&&\\
&\multicolumn{2}{l}{$n=35$}&&  &  &&&&  &  &&&&\\
0.00 & 4.5 & 5.2 & 6.4 & 6.3 & 6.5 & 4.4 & 4.4 & 3.8 & 4.1 & 5.0 & 5.6 & 5.4
& 3.0 & 4.1 & 5.3 \\
0.25 & 4.4 & 2.4 & 5.2 & 2.6 & 4.8 & 4.4 & 4.6 & 4.0 & 4.2 & 4.3 & 4.8 & 4.8
& 2.7 & 4.2 & 4.4 \\
0.50 & 5.2 & 0.9 & 7.0 & 1.2 & 6.6 & 5.5 & 5.3 & 4.9 & 5.2 & 5.8 & 6.0 & 5.8 &
3.3 & 5.2 & 5.3 \\
0.75 & 4.4 & 0.0 & 4.9 & 0.1 & 5.5 & 5.4 & 4.8 & 4.4 & 4.8 & 5.1 & 4.2 & 4.0 &
2.6 & 3.4 & 3.9 \\
\end{tabular}
%\end{scriptsize}
\end{table}

\clearpage

{\small\tabcolsep=5.5pt
\begin{table}[!t]
\caption[Empirical sizes and powers of all post hoc tests obtained in model~M2 under normal distribution]{Empirical sizes (1-2) and powers (1-3, 2-3) (as percentages) of all post hoc tests obtained in model~M2 under normal distribution.}
\label{table9_sm}
\centering
\begin{footnotesize}
%\begin{scriptsize}
\begin{tabular}{ll rrrrr rrrrr rrrrr}%
\hline\noalign{\smallskip}
&&\multicolumn{5}{l}{$\mathcal{C}_n(3)$}  & \multicolumn{5}{l}{$\mathcal{D}_n3)$}  & \multicolumn{5}{l}{$\mathcal{E}_n(3)$}\\%\cline{2-6}\cline{7-11}\cline{12-16}
$\rho$&Comp &$\mathcal{P}_1$&$\mathcal{P}_2$&$\mathcal{B}_1$&$\mathcal{B}_2$&$\mathcal{B}_3$  & $\mathcal{P}_1$&$\mathcal{P}_2$&$\mathcal{B}_1$&$\mathcal{B}_2$&$\mathcal{B}_3$  &  $\mathcal{P}_1$&$\mathcal{P}_2$&$\mathcal{B}_1$&$\mathcal{B}_2$&$\mathcal{B}_3$\\
\hline\noalign{\smallskip}
$n=35$ & & & & & & & & & & & & & & & & \\
0.00 & 1-2 & 2.1 & 1.7 & 2.2 & 2.1 & 2.1 & 2.1 & 1.9 & 2.0 & 1.9 & 2.0 &
1.3 & 1.6 & 0.9 & 1.4 & 1.5 \\
& 1-3 & 12.8 & 12.2 & 12.7 & 13.1 & 12.6 & 17.1 & 16.6 & 12.1 & 14.1 &
15.2 & 62.9 & 62.3 & 55.9 & 59.8 & 63.8 \\
& 2-3 & 14.7 & 14.7 & 15.0 & 14.7 & 14.5 & 20.7 & 19.3 & 15.9 & 16.4 &
18.7 & 63.5 & 62.7 & 56.0 & 60.1 & 64.5 \\
0.25 & 1-2 & 2.3 & 0.4 & 2.7 & 0.6 & 2.3 & 1.9 & 2.4 & 2.0 & 1.8 & 2.2 &
1.8 & 1.7 & 0.7 & 1.5 & 2.1 \\
& 1-3 & 16.2 & 11.9 & 16.3 & 12.1 & 16.5 & 22.8 & 20.9 & 16.0 & 17.5 &
18.9 & 65.5 & 66.7 & 59.5 & 63.2 & 67.6 \\
\end{tabular}
%\end{scriptsize}
\end{footnotesize}
\end{table}
}

\clearpage

{\small\tabcolsep=5.5pt
\begin{table}[!t]
\caption[Empirical sizes and powers of all post hoc tests obtained in model~M2 under lognormal distribution]{Empirical sizes (1-2) and powers (1-3, 2-3) (as percentages) of all post hoc tests obtained in model~M2 under lognormal distribution.}
\label{table10_sm}
\centering
\begin{footnotesize}
%\begin{scriptsize}
\begin{tabular}{ll rrrrr rrrrr rrrrr}%
\hline\noalign{\smallskip}
&&\multicolumn{5}{l}{$\mathcal{C}_n(3)$}  & \multicolumn{5}{l}{$\mathcal{D}_n3)$}  & \multicolumn{5}{l}{$\mathcal{E}_n(3)$}\\%\cline{2-6}\cline{7-11}\cline{12-16}
$\rho$&Comp &$\mathcal{P}_1$&$\mathcal{P}_2$&$\mathcal{B}_1$&$\mathcal{B}_2$&$\mathcal{B}_3$  & $\mathcal{P}_1$&$\mathcal{P}_2$&$\mathcal{B}_1$&$\mathcal{B}_2$&$\mathcal{B}_3$  &  $\mathcal{P}_1$&$\mathcal{P}_2$&$\mathcal{B}_1$&$\mathcal{B}_2$&$\mathcal{B}_3$\\
\hline\noalign{\smallskip}
$n=35$ & & & & & & & & & & & & & & & & \\
0.00 & 1-2 & 2.5 & 2.3 & 2.5 & 2.5 & 2.5 & 2.5 & 2.2 & 2.2 & 2.0 & 2.3 &
2.5 & 3.0 & 1.4 & 1.9 & 2.5 \\
& 1-3 & 64.5 & 66.7 & 61.0 & 64.5 & 60.7 & 63.2 & 62.8 & 53.7 & 57.3 &
59.0 & 78.6 & 79.2 & 69.0 & 73.6 & 78.7 \\
& 2-3 & 65.6 & 64.5 & 62.1 & 63.2 & 62.0 & 64.1 & 63.1 & 55.3 & 58.1 &
59.3 & 80.1 & 80.0 & 70.2 & 74.7 & 79.8 \\
0.25 & 1-2 & 1.5 & 0.2 & 1.5 & 0.2 & 1.2 & 1.4 & 1.3 & 1.0 & 0.9 & 1.0 &
1.3 & 1.2 & 0.9 & 0.9 & 1.1 \\
& 1-3 & 68.5 & 63.5 & 63.8 & 60.1 & 63.1 & 67.0 & 66.8 & 56.6 & 59.4 &
61.2 & 82.9 & 82.5 & 72.5 & 77.9 & 83.8 \\
\end{tabular}
%\end{scriptsize}
\end{footnotesize}
\end{table}
}

\clearpage

{\small\tabcolsep=5.5pt
\begin{table}[!t]
\caption[Empirical sizes and powers of all post hoc tests obtained in model~M3 under normal distribution]{Empirical sizes (1-2) and powers (1-3, 2-3) (as percentages) of all post hoc tests obtained in model~M3 under normal distribution.}
\label{table11_sm}
\centering
\begin{footnotesize}
%\begin{scriptsize}
\begin{tabular}{ll rrrrr rrrrr rrrrr}%
\hline\noalign{\smallskip}
&&\multicolumn{5}{l}{$\mathcal{C}_n(3)$}  & \multicolumn{5}{l}{$\mathcal{D}_n3)$}  & \multicolumn{5}{l}{$\mathcal{E}_n(3)$}\\%\cline{2-6}\cline{7-11}\cline{12-16}
$\rho$&Comp &$\mathcal{P}_1$&$\mathcal{P}_2$&$\mathcal{B}_1$&$\mathcal{B}_2$&$\mathcal{B}_3$  & $\mathcal{P}_1$&$\mathcal{P}_2$&$\mathcal{B}_1$&$\mathcal{B}_2$&$\mathcal{B}_3$  &  $\mathcal{P}_1$&$\mathcal{P}_2$&$\mathcal{B}_1$&$\mathcal{B}_2$&$\mathcal{B}_3$\\
\hline\noalign{\smallskip}
$n=35$ & & & & & & & & & & & & & & & & \\
0.00 & 1-2 & 1.6 & 1.7 & 2.1 & 1.9 & 2.3 & 1.4 & 1.3 & 1.3 & 1.4 & 1.7 &
0.7 & 0.9 & 0.4 & 0.6 & 1.0 \\
& 1-3 & 70.8 & 70.2 & 67.8 & 69.7 & 67.4 & 89.8 & 88.7 & 82.8 & 84.0 &
86.0 & 99.8 & 99.7 & 99.3 & 99.6 & 99.9 \\
& 2-3 & 71.2 & 71.7 & 70.0 & 68.9 & 68.8 & 89.3 & 89.0 & 83.2 & 84.7 &
85.6 & 99.8 & 99.8 & 99.3 & 99.7 & 99.8 \\
0.25 & 1-2 & 2.1 & 0.4 & 2.3 & 0.7 & 2.4 & 1.7 & 1.7 & 1.7 & 1.7 & 1.6 &
1.6 & 1.5 & 0.9 & 1.2 & 1.6 \\
& 1-3 & 77.4 & 71.6 & 74.9 & 69.4 & 73.8 & 94.5 & 94.1 & 89.6 & 90.8 &
92.0 & 100 & 99.9 & 99.8 & 99.9 & 99.9 \\
\end{tabular}
%\end{scriptsize}
\end{footnotesize}
\end{table}
}

\clearpage

{\small\tabcolsep=5.5pt
\begin{table}[!t]
\caption[Empirical sizes and powers of all post hoc tests obtained in model~M3 under lognormal distribution]{Empirical sizes (1-2) and powers (1-3, 2-3) (as percentages) of all post hoc tests obtained in model~M3 under lognormal distribution.}
\label{table12_sm}
\centering
\begin{footnotesize}
%\begin{scriptsize}
\begin{tabular}{ll rrrrr rrrrr rrrrr}%
\hline\noalign{\smallskip}
&&\multicolumn{5}{l}{$\mathcal{C}_n(3)$}  & \multicolumn{5}{l}{$\mathcal{D}_n3)$}  & \multicolumn{5}{l}{$\mathcal{E}_n(3)$}\\%\cline{2-6}\cline{7-11}\cline{12-16}
$\rho$&Comp &$\mathcal{P}_1$&$\mathcal{P}_2$&$\mathcal{B}_1$&$\mathcal{B}_2$&$\mathcal{B}_3$  & $\mathcal{P}_1$&$\mathcal{P}_2$&$\mathcal{B}_1$&$\mathcal{B}_2$&$\mathcal{B}_3$  &  $\mathcal{P}_1$&$\mathcal{P}_2$&$\mathcal{B}_1$&$\mathcal{B}_2$&$\mathcal{B}_3$\\
\hline\noalign{\smallskip}
$n=35$ & & & & & & & & & & & & & & & & \\
0.00 & 1-2 & 2.3 & 2.1 & 2.2 & 2.0 & 2.3 & 2.6 & 2.5 & 1.5 & 2.1 & 2.2 &
2.2 & 2.2 & 1.2 & 1.9 & 2.0 \\
& 1-3 & 99.9 & 100 & 99.9 & 99.9 & 99.8 & 99.8 & 99.8 & 99.6 & 99.5 &
99.6 & 99.8 & 99.8 & 99.7 & 99.8 & 99.7 \\
& 2-3 & 99.9 & 99.9 & 99.8 & 99.9 & 99.8 & 99.8 & 99.8 & 99.6 & 99.7 &
99.8 & 100 & 100 & 99.9 & 99.8 & 100 \\
0.25 & 1-2 & 1.2 & 0.3 & 1.1 & 0.3 & 1.3 & 1.4 & 1.3 & 1.0 & 1.0 & 1.2 &
1.7 & 1.4 & 0.7 & 1.3 & 1.8 \\
& 1-3 & 100 & 99.9 & 99.8 & 99.9 & 99.9 & 99.9 & 99.9 & 99.7 & 99.8 &
99.9 & 99.9 & 99.9 & 99.7 & 99.9 & 99.9 \\
\end{tabular}
%\end{scriptsize}
\end{footnotesize}
\end{table}
}

\clearpage

{\small\tabcolsep=5.5pt
\begin{table}[!t]
\caption[Empirical sizes and powers of all post hoc tests obtained in model~M5 under normal distribution]{Empirical sizes (1-2) and powers (1-3, 2-3) (as percentages) of all post hoc tests obtained in model~M5 under normal distribution.}
\label{table13_sm}
\centering
\begin{footnotesize}
%\begin{scriptsize}
\begin{tabular}{ll rrrrr rrrrr rrrrr}%
\hline\noalign{\smallskip}
&&\multicolumn{5}{l}{$\mathcal{C}_n(3)$}  & \multicolumn{5}{l}{$\mathcal{D}_n3)$}  & \multicolumn{5}{l}{$\mathcal{E}_n(3)$}\\%\cline{2-6}\cline{7-11}\cline{12-16}
$\rho$&Comp &$\mathcal{P}_1$&$\mathcal{P}_2$&$\mathcal{B}_1$&$\mathcal{B}_2$&$\mathcal{B}_3$  & $\mathcal{P}_1$&$\mathcal{P}_2$&$\mathcal{B}_1$&$\mathcal{B}_2$&$\mathcal{B}_3$  &  $\mathcal{P}_1$&$\mathcal{P}_2$&$\mathcal{B}_1$&$\mathcal{B}_2$&$\mathcal{B}_3$\\
\hline\noalign{\smallskip}
$n=35$ & & & & & & & & & & & & & & & & \\
0.00 & 1-2 & 1.2 & 1.0 & 1.4 & 1.4 & 1.6 & 1.0 & 1.0 & 0.9 & 1.0 & 1.3 &
1.2 & 1.4 & 0.6 & 0.9 & 1.1 \\
& 1-3 & 32.4 & 33.7 & 35.3 & 35.1 & 34.6 & 36.8 & 37.7 & 33.1 & 33.9 &
36.6 & 66.3 & 66.1 & 57.2 & 62.3 & 66.5 \\
& 2-3 & 33.0 & 33.9 & 36.3 & 36.2 & 35.3 & 37.3 & 37.1 & 33.8 & 34.2 &
36.7 & 68.1 & 67.3 & 60.5 & 64.6 & 68.2 \\
0.25 & 1-2 & 1.4 & 0.7 & 2.1 & 0.8 & 1.9 & 1.4 & 1.7 & 1.3 & 1.3 & 1.6 &
1.3 & 1.5 & 0.7 & 1.2 & 1.3 \\
& 1-3 & 36.1 & 34.7 & 39.9 & 35.3 & 38.5 & 41.0 & 40.5 & 37.1 & 38.8 &
39.5 & 69.6 & 69.5 & 61.6 & 66.8 & 69.5 \\
\end{tabular}
%\end{scriptsize}
\end{footnotesize}
\end{table}
}

\newpage

{\small\tabcolsep=5.5pt
\begin{table}[!t]
\caption[Empirical sizes and powers of all post hoc tests obtained in model~M5 under lognormal distribution]{Empirical sizes (1-2) and powers (1-3, 2-3) (as percentages) of all post hoc tests obtained in model~M5 under lognormal distribution.}
\label{table14_sm}
\centering
\begin{footnotesize}
%\begin{scriptsize}
\begin{tabular}{ll rrrrr rrrrr rrrrr}%
\hline\noalign{\smallskip}
&&\multicolumn{5}{l}{$\mathcal{C}_n(3)$}  & \multicolumn{5}{l}{$\mathcal{D}_n3)$}  & \multicolumn{5}{l}{$\mathcal{E}_n(3)$}\\%\cline{2-6}\cline{7-11}\cline{12-16}
$\rho$&Comp &$\mathcal{P}_1$&$\mathcal{P}_2$&$\mathcal{B}_1$&$\mathcal{B}_2$&$\mathcal{B}_3$  & $\mathcal{P}_1$&$\mathcal{P}_2$&$\mathcal{B}_1$&$\mathcal{B}_2$&$\mathcal{B}_3$  &  $\mathcal{P}_1$&$\mathcal{P}_2$&$\mathcal{B}_1$&$\mathcal{B}_2$&$\mathcal{B}_3$\\
\hline\noalign{\smallskip}
$n=35$ & & & & & & & & & & & & & & & & \\
0.00 & 1-2 & 1.7 & 2.1 & 2.1 & 1.8 & 1.9 & 2.2 & 1.9 & 1.6 & 1.5 & 2.0 &
1.2 & 1.2 & 0.8 & 1.1 & 1.2 \\
& 1-3 & 100 & 100 & 99.8 & 100 & 99.7 & 99.9 & 99.8 & 98.4 & 99.4 & 99.6
& 89.9 & 90.0 & 84.1 & 87.7 & 91.3 \\
& 2-3 & 99.9 & 100 & 99.7 & 99.8 & 99.7 & 99.8 & 99.8 & 98.9 & 98.8 &
99.2 & 89.4 & 89.5 & 82.9 & 86.6 & 89.6 \\
0.25 & 1-2 & 1.8 & 0.5 & 1.4 & 0.2 & 1.7 & 1.5 & 1.6 & 1.1 & 1.5 & 1.6 &
1.4 & 1.6 & 0.9 & 1.1 & 1.4 \\
& 1-3 & 99.9 & 100 & 99.8 & 99.9 & 99.9 & 99.9 & 99.9 & 99.4 & 99.7 &
99.6 & 92.6 & 92.4 & 87.8 & 91.0 & 93.5 \\
\end{tabular}
%\end{scriptsize}
\end{footnotesize}
\end{table}
}

\newpage

{\small\tabcolsep=5.5pt
\begin{table}[!t]
\caption[Empirical sizes and powers of all post hoc tests obtained in model~M6 under normal distribution]{Empirical sizes (1-2) and powers (1-3, 2-3) (as percentages) of all post hoc tests obtained in model~M6 under normal distribution.}
\label{table15_sm}
\centering
\begin{footnotesize}
%\begin{scriptsize}
\begin{tabular}{ll rrrrr rrrrr rrrrr}%
\hline\noalign{\smallskip}
&&\multicolumn{5}{l}{$\mathcal{C}_n(3)$}  & \multicolumn{5}{l}{$\mathcal{D}_n3)$}  & \multicolumn{5}{l}{$\mathcal{E}_n(3)$}\\%\cline{2-6}\cline{7-11}\cline{12-16}
$\rho$&Comp &$\mathcal{P}_1$&$\mathcal{P}_2$&$\mathcal{B}_1$&$\mathcal{B}_2$&$\mathcal{B}_3$  & $\mathcal{P}_1$&$\mathcal{P}_2$&$\mathcal{B}_1$&$\mathcal{B}_2$&$\mathcal{B}_3$  &  $\mathcal{P}_1$&$\mathcal{P}_2$&$\mathcal{B}_1$&$\mathcal{B}_2$&$\mathcal{B}_3$\\
\hline\noalign{\smallskip}
$n=35$ & & & & & & & & & & & & & & & & \\
0.00 & 1-2 & 1.5 & 1.6 & 2.1 & 1.6 & 2.3 & 1.7 & 1.9 & 1.8 & 1.7 & 1.9 &
2.2 & 2.1 & 1.2 & 1.8 & 2.0 \\
& 1-3 & 44.4 & 45.0 & 47.1 & 47.0 & 47.0 & 51.5 & 51.1 & 46.5 & 48.7 &
49.8 & 76.1 & 75.8 & 69.4 & 73.6 & 77.0 \\
& 2-3 & 42.9 & 45.7 & 47.7 & 47.6 & 48.1 & 49.3 & 50.9 & 46.0 & 48.0 &
50.0 & 75.7 & 75.5 & 67.6 & 71.7 & 76.7 \\
0.25 & 1-2 & 2.3 & 1.0 & 2.9 & 1.0 & 3.0 & 2.1 & 2.2 & 1.8 & 2.2 & 2.2 &
2.3 & 2.0 & 1.3 & 2.1 & 1.9 \\
& 1-3 & 47.2 & 44.7 & 51.0 & 46.6 & 51.0 & 54.3 & 54.2 & 50.9 & 52.2 &
54.9 & 79.1 & 79.4 & 70.3 & 75.3 & 79.6 \\
\end{tabular}
%\end{scriptsize}
\end{footnotesize}
\end{table}
}

\newpage

{\small\tabcolsep=5.5pt
\begin{table}[!t]
\caption[Empirical sizes and powers of all post hoc tests obtained in model~M6 under lognormal distribution]{Empirical sizes (1-2) and powers (1-3, 2-3) (as percentages) of all post hoc tests obtained in model~M6 under lognormal distribution.}
\label{table16_sm}
\centering
\begin{footnotesize}
%\begin{scriptsize}
\begin{tabular}{ll rrrrr rrrrr rrrrr}%
\hline\noalign{\smallskip}
&&\multicolumn{5}{l}{$\mathcal{C}_n(3)$}  & \multicolumn{5}{l}{$\mathcal{D}_n3)$}  & \multicolumn{5}{l}{$\mathcal{E}_n(3)$}\\%\cline{2-6}\cline{7-11}\cline{12-16}
$\rho$&Comp &$\mathcal{P}_1$&$\mathcal{P}_2$&$\mathcal{B}_1$&$\mathcal{B}_2$&$\mathcal{B}_3$  & $\mathcal{P}_1$&$\mathcal{P}_2$&$\mathcal{B}_1$&$\mathcal{B}_2$&$\mathcal{B}_3$  &  $\mathcal{P}_1$&$\mathcal{P}_2$&$\mathcal{B}_1$&$\mathcal{B}_2$&$\mathcal{B}_3$\\
\hline\noalign{\smallskip}
$n=35$ & & & & & & & & & & & & & & & & \\
0.00 & 1-2 & 1.6 & 2.1 & 1.5 & 2.1 & 1.5 & 1.9 & 2.0 & 1.2 & 1.6 & 1.5 &
2.1 & 2.3 & 1.3 & 1.6 & 2.4 \\
& 1-3 & 100 & 100 & 100 & 100 & 100 & 100 & 100 & 100 & 100 & 100 & 97.4
& 96.8 & 93.6 & 96.1 & 97.3 \\
& 2-3 & 100 & 100 & 100 & 100 & 100 & 100 & 100 & 100 & 100 & 100 & 97.7
& 97.0 & 94.3 & 95.5 & 97.2 \\
0.25 & 1-2 & 2.3 & 0.4 & 2.2 & 0.4 & 2.2 & 1.9 & 1.9 & 1.6 & 1.4 & 1.8 &
2.5 & 2.7 & 1.6 & 2.0 & 2.5 \\
& 1-3 & 100 & 100 & 100 & 100 & 100 & 100 & 100 & 100 & 100 & 100 & 98.0
& 97.9 & 95.6 & 96.6 & 97.9 \\
\end{tabular}
%\end{scriptsize}
\end{footnotesize}
\end{table}
}

\end{document}